\documentclass[prb,amsmath,amssymb,superscriptaddress,twocolumn]{revtex4-1}

\usepackage{graphicx}
\usepackage{dcolumn}
\usepackage{epsfig}
\usepackage{xcolor}

\usepackage[bbgreekl]{mathbbol}

\usepackage{mathrsfs}
\usepackage{eufrak}

\newcommand\vex[1]{\mathbf{#1}}
\newcommand\gvex[1]{\boldsymbol{#1}}

\def\bra#1{\mathinner{\langle{#1}|}}
\def\ket#1{\mathinner{|{#1}\rangle}}
\def\inner#1#2{\mathinner{\langle{#1}|{#2}\rangle}}

\def\bbra#1{\mathinner{\langle\hspace{-0.75mm}\langle{#1}|}}
\def\kket#1{\mathinner{|{#1}\rangle\hspace{-0.75mm}\rangle}}

\def\sgn{\mathrm{sgn}}
\def\re{\mathrm{Re}\,}
\def\im{\mathrm{Im}\,}
\def\Tr{\mathrm{Tr}}
\def\tr{\mathrm{tr}}

\def\id{\mathbb{1}} 

\def\Texp{\mathrm{T}\hspace{-1mm}\exp}

\def\nodag{^{\vphantom{\dag}}}

\usepackage[T3,T1]{fontenc}
\DeclareSymbolFont{tipa}{T3}{cmr}{m}{n}
\DeclareMathAccent{\invbreve}{\mathalpha}{tipa}{16}

\usepackage{accents}
\newlength{\hhatheight}
\newcommand{\hhat}[1]{%
    \settoheight{\hhatheight}{\ensuremath{\hat{#1}}}%
    \addtolength{\hhatheight}{-0.35ex}%
    \hat{\vphantom{\rule{1pt}{\hhatheight}}%
    \smash{\hat{#1}}}
}

\usepackage{xcolor}
\newcommand{\extra}[1]{}

\makeatother

\begin{document}

\title{Linear Response Theory and Optical Conductivity of Floquet Topological Insulators}

\author{Abhishek Kumar}
\affiliation{Department of Physics, Indiana University, Bloomington, Indiana 47405,
USA}

\author{M. Rodriguez-Vega}
\affiliation{Department of Physics, The University of Texas at Austin, Austin, TX 78712, USA}
\affiliation{Department of Physics, Northeastern University, Boston, MA 02115, USA}

\author{T. Pereg-Barnea}
\affiliation{Department of Physics, McGill University, Montr\'eal, Qu\'ebec H3A 2T8, Canada}

\author{B. Seradjeh}
\affiliation{Department of Physics, Indiana University, Bloomington, Indiana 47405,
USA}

\begin{abstract}
Motivated by the quest for experimentally accessible dynamical probes of Floquet topological insulators, we formulate the linear response theory of a periodically driven system. We illustrate the applications of this formalism by giving general expressions for optical conductivity of Floquet systems, including its homodyne and heterodyne components and beyond. We obtain the Floquet optical conductivity of specific driven models, including two-dimensional Dirac material such as the surface of a topological insulator, graphene, and the Haldane model irradiated with circularly or linearly polarized laser, as well as semiconductor quantum well driven by an ac potential. We obtain approximate analytical expressions and perform numerically exact calculations of the Floquet optical conductivity in  different scenarios of the occupation of the Floquet bands, in particular the diagonal Floquet distribution and the distribution obtained after a quench. We comment on experimental signatures and detection of Floquet topological phases using optical probes.
\end{abstract}

\date{\today}

{
\let\clearpage\relax
\maketitle
}

\section{Introduction}
Optical properties of solids are widely studied, both experimentally and theoretically~\cite{Dresselhaus_2018}. As a result of its interaction with light, quantum properties of matter, such as its conductivity, can be modified, thus enabling optically activated devices~\cite{drever1983,Tame_2013,Zhang_2014,He_2015,SieLuiLee17a}. Beside such device applications, the optical response of material provides a powerful way to probe the quantum states of electrons and their excitations in spatially periodic potentials. This response is well understood when the light intensity is sufficiently weak to be treated perturbatively within the framework of linear response theory. In particular, the Kubo formalism connects the equilibrium electronic band structure to various linear response coefficients, such as optical conductivity~\cite{kubo1957}.

Recently, the effects of strong light-matter interaction have come to the forefront of materials research. In particular, it was understood theoretically using the Floquet theory of periodically driven systems, and shown experimentally, that the electronic bands in the material can themselves be modified under the intense field of light. In this way, new electronic phases, such as quantum Hall states in graphene, have been proposed to be produced dynamically out of equilibrium~\cite{Oka,wu2011,Gu,Fregoso2013,Lindner,Kitagawa,Kundu2,Usaj}. For example, time-resolved spectroscopy of surfaces of topological insulators under intense laser has shown modified Floquet-Bloch bands consistent with the predictions of Floquet theory~\cite{Wang}. 

The modification of bands leaves a trace in a variety of physical properties, such as transport coefficients~\cite{Kundu,Kitagawa,Gu,Oka,Dehghani2,Dehghani3,Farrell,Farrell2,chen2018,Perez-Piskunow2014,Torres2014,caio2016,Higuchi2017,wackerl2019floquetdrude} and quantum noise~\cite{Moskalets,Rodriguez-Vega,Rodriguez-Vega2}, thus allowing its detection. However, spectroscopic measurements are constrained in what they can probe or how readily they may be set up while the system is externally driven. Thus, it is desirable to have a larger toolbox of probes of the Floquet-Bloch bands. In this toolbox, the optical response of the system stands out since it can detect not only the dynamical modifications of the bands, but also their bulk topology, for example in the DC limit of the optical Hall conductivity~\cite{Rechtsman,McIver2020,Dehghani2}.

In this paper, we extend the equilibrium linear response theory to the Floquet theory of a strongly driven system out of equilibrium, which is probed by a weak external potential. We show that even in this linear-probe regime, the strong driving of the system results in a response not only at the frequency of the probe, but at all its harmonic displacements by the frequency of the drive. Thus, probing a system driven at frequency $\Omega$ at a probe frequency $\omega$ produces a signal at $\omega+n\Omega$, $n\in\mathbb{Z}$. Our theory greatly expands the existing literature~\cite{Dehghani3,Du} by naturally incorporating not just the homodyne ($n=0$) but also the heterodyne ($n=\pm 1$) response~\cite{lenth1983,maznev1998,oka2016} as well as all higher harmonics of the drive.

We provide a general expression for Floquet optical conductivity tensor, elucidate its optical sum rules, and present analytical expressions in the DC limit. We illustrate the structure of this optical response in specific models through analytic and numerical calculations. We recover previous results for quantization of the homodyne optical Hall conductivity in the DC limit in terms of the Chern number of occupied Floquet-Bloch bands~\cite{Dehghani2}. Moreover, we show that the heterodyne optical conductivity also detects the Floquet topological transitions in the DC limit as a singular enhancement at the transition. 

The paper is organized as follows. In Sec.~\ref{sec:FloqLR}, we present the general theory of linear response of a periodically driven system using Floquet theory. In Sec.~\ref{sec:FloqOC}, we use this theory to give a general expression for the Floquet optical conductivity and obtain its DC limit. In Sec.~\ref{sec:analytic}, we present analytical expressions in the high-frequency limit for the optical conductivity of a general two-band model and illustrate its utility for a driven Dirac cone. In Sec.~\ref{sec:numerics}, we present numerical solutions for the homodyne and heterodyne optical conductivity tensor of two-dimensional periodically driven models~\cite{oka2016}. We conclude in Sec.~\ref{sec:discuss} with a discussion. Some details of our calculations are given in the Appendix.

\vspace{-3mm}
\section{Linear Response in Floquet Formalism}\label{sec:FloqLR}
\vspace{-2mm}
\subsection{Primer on Floquet Theory}
\vspace{-2mm}
Using the Floquet theorem for a periodic Hamiltonian $\hat{H}(t) = \hat{H}(t+T)$ with period $T=2\pi/\Omega$, the evolution operator $\hat{U}(t,t_0) = \Texp{\left[-i\int_{t_0}^t \hat H(s) ds \right]}$, where $\Texp$ is the time-ordered exponential, can be decomposed as~\cite{Flo83a,Sam73a}
\begin{align}
\hat{U}(t,t_0) = e^{-i(t-t_0)\hat H_F(t)}\hat \Phi(t,t_0),
\end{align}
into a periodic micromotion operator
\begin{align}
\hat \Phi(t,t_0) &= \hat \Phi(t+T,t_0) = \hat \Phi(t,t_0+T) \nonumber \\
&\equiv \sum_\alpha \ket{\phi_\alpha(t)}\bra{\phi_\alpha(t_0)},
\end{align}
and the evolution under the Floquet Hamiltonian
\begin{align}
\hat H_F(t) = \sum_\alpha \epsilon_\alpha \ket{\phi_\alpha(t)}\bra{\phi_\alpha(t)},
\end{align}
with quasienergy eigenvalues $\epsilon_\alpha$ (independent of $t$), both written in the basis of periodic Floquet states $\ket{\phi_\alpha(t)} = \ket{\phi_\alpha(t+T)}$ that are solutions of the Floquet-Schr\"odinger equation,
\begin{equation}
    [\hat H(t) - i\partial_t]\ket{\phi_\alpha(t)} = \epsilon_\alpha \ket{\phi_\alpha(t)}.
\end{equation}

This structure can be formalized in the extended Floquet-Hilbert space~\cite{Shi65a,EckAni15a,Rodriguez-Vega3} $\mathscr{F}=\mathscr{H}\otimes\mathscr{I}$, where $\mathscr{H}$ is the usual Hilbert space and $\mathscr{I}$ is the auxiliary space of periodic functions spanned by an orthonormal basis $|t) = |t+T)$ with 
\begin{equation}
(t'|t)=\breve \delta(t-t')\equiv\sum_{p\in\mathbb{Z}}\delta(t-t'-pT)
\end{equation}
and $\int_0^T |t)(t| dt/T = \breve{I}$, the identity oeprator in $\mathscr{I}$. 
Equivalently, we may define an orthonormal Fourier basis 
\begin{equation}
|n) = \int_0^T e^{-in\Omega t}|t)dt/T
\end{equation}
for $n\in\mathbb{Z}$ with $(n'|n) = \delta_{nn'}$, $\sum_n |n)(n| = \breve I$. A periodic state $\ket{\phi(t)}\in\mathscr{H}$ can be ``lifted'' to 
\begin{equation}
\kket{\phi_t} \equiv \ket{\phi(t)}\otimes|t)\in\mathscr{F}. 
\end{equation}
We also define a set of Fourier states 
\begin{equation}
\kket{\phi_m} = \sum_{n}\ket{\phi^{(n+m)}}\otimes|n)\in\mathscr{F}, 
\end{equation}
where the Fourier components 
\begin{equation}
\ket{\phi^{(n)}} = \int_0^T e^{in\Omega t}\ket{\phi(t)} dt/T.
\end{equation} 
Then, $\kket{\phi_t} = \sum_m e^{-im\Omega t}\kket{\phi_m}$. 

For a periodic operator $\hat H(t)=\hat H(t+T)$ acting on $\mathscr{H}$, we define
\begin{equation}
\hhat{H} = \int_0^T \hat H(t)\otimes|t)(t| \frac{dt}T = \sum_{n,m} \hat H^{(n-m)} \otimes |n)(m|,    
\end{equation} 
acting on $\mathscr{F}$ with Fourier components 
\begin{equation}
\hat H^{(n)} = \int_0^T e^{in\Omega t} \hat H(t) dt/T.
\end{equation}
The Fourier shift operator $\hhat{\mu}_n\kket{\phi_m} = \kket{\phi_{n+m}}$ is given by 
\begin{equation}
\hhat{\mu}_n = \mathbb{1}\otimes \int_0^T |t)e^{in\Omega t}(t| dt/T.
\end{equation} 
We also define a time-derivative operator 
\begin{equation}
\hhat{Z}_t = \mathbb{1}\otimes \sum_n |n) n\Omega (n|,
\end{equation}
with the action
\begin{align}
\hhat Z_t \kket{\phi_m} 
	&= i\kket{(d\phi/d t)_m} \nonumber \\
	& \equiv\int_0^T e^{im\Omega t} [id\ket{\phi(t)}/dt]\otimes|t) dt/T.
\end{align}
The Floquet-Schr\"odinger equation in this extended space takes the form 
\begin{equation}
    (\hhat H - \hhat Z_t)\kket{\phi_{\alpha m}} = \epsilon_{\alpha m} \kket{\phi_{\alpha m}},
\end{equation}
where $\kket{\phi_{\alpha m}}$
form a complete basis for $\mathscr{F}$, and $\epsilon_{\alpha m} = \epsilon_\alpha + m\Omega$ with quasienergies  $\epsilon_\alpha$.

Finally, we note that in the extended Floquet-Hilbert space, we can represent the Floquet Green's function
\begin{equation}\label{eq:FG}
\hhat{G}(\omega) = (\omega - \hhat{H} + \hhat{Z}_t)^{-1} = \lim_{\eta\to0^+}\sum_{\alpha m}\frac{\kket{\phi_{\alpha m}}\bbra{\phi_{\alpha m}}}{\omega - \epsilon_\alpha - m\Omega + i\eta}.
\end{equation}
Back in the Hilbert space $\mathscr{H}$, we have the Fourier components
\begin{align}
    \hat G^{(n)}(\omega) 
    &\equiv (0|\hhat G(\omega) |n) \nonumber\\
    &= \lim_{\eta\to0^+}\sum_{\alpha m}\frac{\ket{\phi_\alpha^{(m)}}\bra{\phi_\alpha^{(m+n)}}}{\omega-\epsilon_\alpha - m\Omega + i\eta}.
    \label{eq:FGn}
\end{align}

\begin{figure}[t]
   \centering
   \includegraphics[width=3.2in]{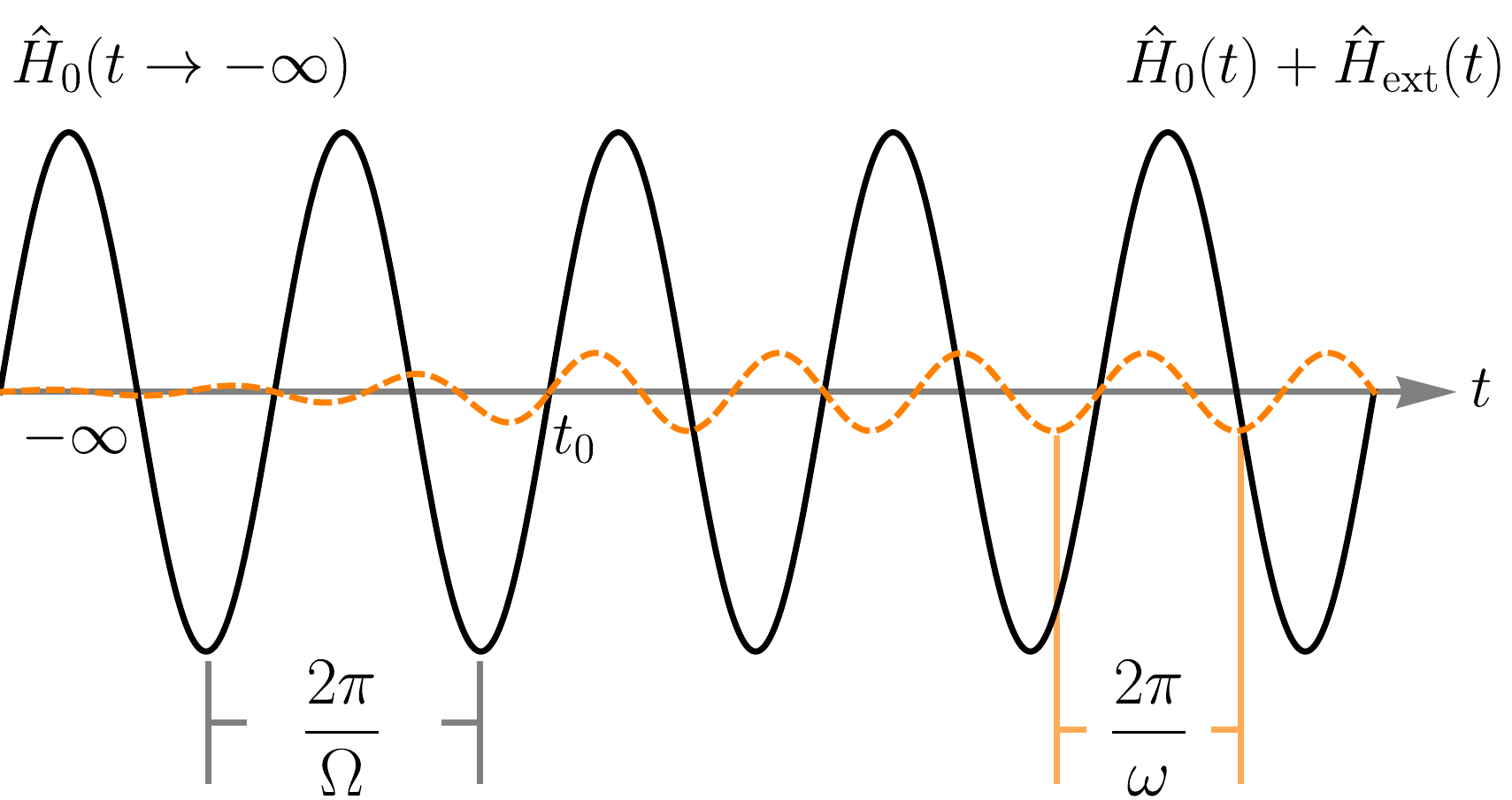}
   \caption{A representation of the component of the dynamical response. Starting at time $t_0$, the response of the periodically driven system with Hamiltonian $\hat{H}_0(t) = \hat H_0(t+2\pi/\Omega)$ at a later time $t>t_0$ is measured in linear order with external perturbation  $\hat{ H}_\text{ext}(t) = \hat{ H}_\text{ext}(t+2\pi/\omega)$. The perturbation is switched on adiabatically long before $t_0$.
   }
   \label{fig:sketch_1}
\end{figure}

\vspace{-10mm}
\subsection{Floquet Linear Response Theory}
\vspace{-2mm}
We formulate the linear response of the driven system in a manner parallel to the linear response theory of an equilibrium system by taking the total Hamiltonian $\hat{{H}}(t) = \hat{{H}}_0(t) + \hat{{H}}_{\text{ext}}(t)$, where $\hat{{H}}_0(t)$ is the periodic Hamiltonian of the driven system and the probe Hamiltonian
\begin{equation}
\hat H_{\text{ext}}(t) = \lim_{\eta\to 0^+} e^{\eta t} \lambda(t) \hat A(t),
\end{equation}
with $\hat A(t)$ the (possibly time-dependent) probe field, is slowly turned on in the distant past, $t\to -\infty$, with the strength $\lambda (t) = \lambda e^{-i\omega t}$ at probe frequency $\omega$, as depicted in Fig.~\ref{fig:sketch_1}. Now we assume a general initial density matrix $\hat\rho_0(t_0)$ (not necessarily thermal) at some initial time $t_0$. Then, the expectation value of an arbitrary operator $\hat B(t)$ is given by ${B(t)} = \Tr[\hat\rho(t)\hat B(t)]$, where $\Tr$ is the trace over the many-body Hilbert space. The change in $B$ due to the external field is
\begin{equation}\label{eq:dBdef}
\delta{B(t)} =  \Tr[\hat\rho(t)\hat B(t)] -  \Tr[\hat\rho_0(t)\hat B_0(t)],
\end{equation}
with the index $0$ indicating the absence of the external field. The linear response is given in terms of the susceptibility 
\vspace{-3mm}
\begin{equation}
\chi_{BA} (t,t') = \left.\frac{\delta{B(t)}}{\delta \lambda(t')}\right\vert_{\lambda=0}.
\end{equation}

In the following, we shall assume all operators can be expanded in a single-particle basis, e.g. the fermionic creation operators $\hat c_{\vex k \mu}^\dagger$ with $\vex k$ the lattice momentum in the first Brillouin zone and $\mu$ some internal degree of freedom (spin, sublattice, band index, etc.), as $\hat H_0(t) = \sum_{\vex k \mu\nu} \hat c_{\vex k \mu}^\dagger [H_0(\vex k, t)]\nodag_{\mu\nu} \hat c\nodag_{\vex k \nu}$. For brevity, we will drop $\vex k$ and treat the operators as matrices in the single-particle basis (diagonal in $\vex k$), with the trace shown by $\tr$. We work in the natural units $\hbar = c = e = 1$.

In order to carry out this calculation, we define an interaction picture via the unperturbed evolution operator $\hat U_0(t,t_0) = \Texp[-i\int_{t_0}^t 
\hat H_0(s) ds]$, so that $\hat \rho^I (t;t_0) =  \hat U_0^\dagger(t,t_0) \hat \rho(t)  \hat U_0(t,t_0)$. Thus,
\begin{equation}
i\frac{\partial}{\partial t}\hat \rho^I(t;t_0) = [\hat H_{\text{ext}}^I(t;t_0),\hat \rho^I(t;t_0)],
\end{equation}
where $\hat H^I_{\text{ext}}(t;t_0) = \hat U_0^\dagger(t,t_0) \hat H_{\text{ext}}(t) \hat U_0(t,t_0)$.
Therefore, to linear order,
\begin{equation}\label{eq:rho1}
\hat \rho^I(t;t_0) = \hat \rho_0(t_0) - i\int_{t_0}^t [\hat H_{\text{ext}}^I(s;t_0),\hat \rho_0(t_0)] ds,
\end{equation}
where we used $\hat \rho^I(t_0;t_0) = \hat \rho_0(t_0)$. Writing the operators such as $\hat B = \sum_{\mu\nu} \hat c^\dagger_\mu[B]_{\mu\nu}\hat c\nodag_\nu$ in the single-particle basis and defining the Green's function $g_{0\nu\mu} = \Tr[{\hat\rho_0(t_0)\hat c^\dagger_{\mu} \hat c\nodag_{\nu}}]$ in Eq.~(\ref{eq:dBdef}), we have
\begin{align}
\delta B(t) 
	= i\int_{t_0}^t \tr\{ g_0 [H^I_\text{ext}(s;t_0),B^I(t;t_0)] \} ds & \nonumber\\
	+ \tr[g_0 \delta B^I(t;t_0)],& \label{eq:dB}
\end{align}
where the single-particle matrices in the interaction picture are defined as $B^I(t;t_0) = U^\dagger(t,t_0) B(t) U(t,t_0)$ with $U(t,t_0) = \Texp[-i\int_{t_0}^t H_0(s) ds]$, and $\delta B^I = B^I - B^I_0$ is the change in the response field itself up to linear order in the probe field. The susceptibility can now be written as
\begin{widetext}
\begin{equation}
\chi_{BA}(t,t';t_0) = \lim_{\eta\to0^+} e^{\eta t'} \bigg[ i \Theta(t-t')\tr\{ g_0 [A^I(t';t_0), B^I(t;t_0) ] \} + \delta(t-t') \tr\{g_0 M^I(t;t_0)\} \bigg],
\end{equation}
\end{widetext}
where $M^I(t;t_0) = \delta B^I(t;t_0)/\delta \lambda(t)|_{\lambda=0}$ and we have assumed $t'>t_0$. 

So far, our development applies to \emph{any} time-dependent Hamiltonian dynamics. Now, we use the Floquet decomposition of the evolution operator $U(t,t_0)$ to write
$
B^I(t;t_0) = \Phi^\dagger(t,t_0)[e^{i(t-t_0)H_F(t)}B(t) e^{-i(t-t_0)H_F(t)}]\Phi(t,t_0).
$
%
Thus,
\begin{align}\label{eq:Bm}
\bra{\phi_\alpha(t_0)} B^I(t;t_0) \ket{\phi_\beta(t_0)} 
	= B^F_{\alpha\beta}(t) e^{-i(t-t_0)(\epsilon_\beta-\epsilon_\alpha)},
\end{align}
with the time-periodic matrix elements,
\begin{align}
B^F_{\alpha\beta}(t) = \bra{\phi_\alpha(t)}B(t)\ket{\phi_\beta(t)} \equiv \sum_m e^{-im\Omega t} B^{F(m)}_{\alpha\beta}.
\end{align}
In the extended Floquet Hilbert space, we have
\begin{equation}
    B_{\alpha\beta}^{F(m-n)} = \bbra{\phi_{\alpha n}} \hhat B \kket{\phi_{\beta m}}.
\end{equation}
Therefore, the matrix elements of the operator in the interaction picture are
\begin{align}\label{eq:FloqBI}
    \bra{\phi_\alpha(t_0)} B^I(t;t_0) \ket{\phi_\beta(t_0)} 
	&= \sum_m \bbra{\phi_{\alpha 0}} \hhat B \kket{\phi_{\beta m}} e^{-im\Omega t} \nonumber\\ &~~~ \times e^{-i(t-t_0)(\epsilon_\beta-\epsilon_\alpha)}.&
\end{align}

The susceptibility can be viewed as a function $\chi_{AB}(t,\tau;\tau_0)$ of $\tau=t-t'$, $\tau_0=t-t_0$, and  a periodic function of $t$. The dependence on $\tau_0$ is a consequence of assuming that the density matrix at the initial time $t_0$ is arbitrary. If this initial matrix is diagonal in the Floquet basis, the dependence on $\tau_0$ will drop away. Alternatively, if we average over this initial time for a fixed initial density matrix, only the diagonal elements of the density matrix will contribute to the susceptibility.
The temporal structure of the susceptibility makes it possible to define its Fourier components,
\begin{equation}
    \chi_{AB}^{(n)}(\omega;\tau_0) = \frac1T\int_0^T dt \int d\tau e^{i n \Omega t + i\omega\tau} \chi_{AB}(t,\tau;\tau_0).
\end{equation}
Using the Floquet matrix elements in Eq.~(\ref{eq:FloqBI}), we find
\begin{widetext}
\begin{align}
    \chi_{AB}^{(n)}(\omega;\tau_0) 
	= \lim_{\eta\to0^+} \sum_{\alpha\beta}g_{0\alpha\beta}
	e^{-i\tau_0(\epsilon_\alpha-\epsilon_\beta+i\eta)}
	\bigg[  
	\sum_{\gamma m} \left(
	\frac{ \bbra{\phi_{\beta -n}} \hhat A \kket{\phi_{\gamma m}} \bbra{\phi_{\gamma m}} \hhat B \kket{\phi_{\alpha 0}}}{\omega + (\epsilon_\alpha -\epsilon_\gamma - m\Omega) + i\eta} \right. & \nonumber\\ \left.
	- \frac{ \bbra{\phi_{\beta 0}} \hhat B \kket{\phi_{\gamma m}} \bbra{\phi_{\gamma m}} \hhat A \kket{\phi_{\alpha n}}}{\omega - (\epsilon_\beta -\epsilon_\gamma - m\Omega) + i\eta} \right) & 
	+ \bbra{\phi_{\beta0}}\hhat M\kket{\phi_{\alpha n}} \bigg].
	\label{eq:chiomega}
\end{align}
We can also write this more compactly using the Floquet Green's function~(\ref{eq:FG}) as
\begin{align}
    \chi_{AB}^{(n)}(\omega;\tau_0) 
	= 
	\sum_{\alpha\beta}g_{0\alpha\beta}
	e^{-i\tau_0(\epsilon_\alpha-\epsilon_\beta+i\eta)}
	\bbra{\phi_{\beta 0}}
	\hhat{A} \hhat{G}(\omega+\epsilon_\alpha+n\Omega) \hhat{B}
	+
	\hhat{B} \hhat{G}^\dagger(-\omega+\epsilon_\beta) \hhat{A}
	+ \hhat M
	\kket{\phi_{\alpha n}}.
\end{align}
\end{widetext}

We note that the reality of $\chi_{AB}(t,\tau;\tau_0)$ imposes the condition $\chi_{AB}^{(-n)}(\omega;\tau_0) = \chi_{AB}^{(n)}(-\omega;\tau_0)^*$. Moreover, since the right hand side of Eq.~(\ref{eq:chiomega}) is an anlytical function of $\omega$ in the upper half of the complex plane (all the residues are in the lower half by virtue of $\eta>0$), the Floquet susceptibility satisfies the Kramers-Kronig relations
\begin{equation}\label{eq:KK}
\tilde \chi^{(n)}_{AB}(\omega) = \frac1{i\pi}\mathcal{P}\int_{-\infty}^{\infty}\frac{\tilde \chi_{AB}^{(n)}(\omega')}{\omega'-\omega}d\omega',
\end{equation}
where $\mathcal{P}\int$ is the principal value of the integral and
\begin{align}
    \tilde\chi_{AB}^{(n)}(\omega) 
    &\equiv \chi_{AB}^{(n)}(\omega) - \chi_{AB}^{(n)}(\infty) \nonumber \\ &
    = \chi_{AB}^{(n)}(\omega) - \sum_{\alpha\beta} g_{0\alpha\beta}e^{-i\tau_0(\epsilon_\alpha-\epsilon_\beta)}\bbra{\phi_{\beta0}}\hhat M\kket{\phi_{\alpha n}}.
\end{align}

\section{Floquet Optical Conductivity}\label{sec:FloqOC}

\subsection{General Expression}
In the specific case of optical conductivity, the probe field is the current and with the strength proportional to the
electromagnetic gauge potential, $\delta \vex A$, which enters the Hamiltonian through minimal coupling $\vex k \to \vex k - \vex A$. Here, $\vex A = \vex A_0+\delta\vex A$ may contain both a drive and the probe fields. The external Hamiltonian has the form $H_{\text{ext}} = - \vex j_0 \cdot \delta \vex A $, where the current 
\begin{equation}
\vex j_0 = %
\vex j\vert_{\vex A\to\vex A_0} = \frac{\partial H_0}{\partial \vex k} \bigg\vert_{\vex k \to \vex k - \vex A_0}.
\end{equation}
The full current operator to linear order is $\vex j = \vex j_0 + \mathbb{m}_0 \delta \vex A$, where the Hermitian matrix 
\begin{equation}
\mathbb{m}_0 = %
\frac{\partial\vex j}{\partial \vex A}\bigg\vert_{\vex A\to\vex A_0} =
\frac{\partial^2 H_0}{\partial \vex k\partial \vex k}\bigg\vert_{\vex k \to \vex k - \vex A_0},
\end{equation}
is the inverse effective mass tensor of the original Hamiltonian, and the term proportional to it is the diamagnetic contribution to current.

The Fourier transform of the current is related to the probe field,
\begin{align}
\delta\vex j(\omega;\tau_0)
    &= \sum_n \bbchi_{jj}^{(n)}(\omega - n\Omega;\tau_0) \delta\vex A(\omega-n\Omega) \nonumber \\
    &= \sum_n \bbsigma^{(n)}(\omega-n\Omega;\tau_0) \delta\vex E(\omega-n\Omega),
\end{align}
%
where $\delta\vex E(\omega) = i\omega \delta\vex A(\omega)$ is the probe electric field, and we have defined the Floquet optical conductivity,
\begin{align}\label{eq:oc}
\bbsigma^{(n)}(\omega;\tau_0) = \frac{\bbchi^{(n)}(\omega;\tau_0)}{i\omega}.
\end{align}
So, unlike the equilibrium response, now the optical current at frequency $\omega$ responds to the field at $\omega-n\Omega$ with the weight $\bbchi^{(n)}(\omega-n\Omega;\tau_0)$ for all $n\in\mathbb{Z}$. For a monochromatic probe field, $\delta\vex E(\omega) = \vex E_{\omega_0} \delta(\omega-\omega_0)$, we have $\delta\vex j(\omega;\tau_0) = [\sum_n \bbsigma^{(n)}(\omega_0;\tau_0) \delta(\omega-\omega_0-n\Omega)]\vex E_{\omega_0}$.
Thus, $\bbsigma^{(n)}(\omega;\tau_0)$ is the component of optical conductivity at frequency $\omega+n\Omega$ in response to the field at $\omega$. As a function of time, we have $\delta\vex j(t)=\bbsigma(\omega_0,t;\tau_0) \delta\vex E_{\omega_0} e^{-i\omega_0t}$ with $\bbsigma(\omega_0,t;\tau_0) = \sum_ne^{-in\Omega t}\bbsigma^{(n)}(\omega_0;\tau_0)$; so, for fixed (or averaged) $\tau_0$, the current shows periodic oscillations with the drive frequency enveloped in the probe frequency.

The occupation of Floquet bands given by $g_{0\alpha\beta}$ is not fixed in our formalism and depends on relaxation processes not considered here~\cite{Dehghani1,Seetharam_2015}. In the following, we assume only the diagonal elements of the density matrix in the Floquet basis contribute to conductivity, either because the density matrix is diagonal, or assuming we average over the initial time. Then,
\begin{widetext}
\begin{align}
\bbsigma^{uv(n)}(\omega) 
	= \frac{i}{\omega} \sum_{\alpha}g_{0\alpha}  \bigg[ \sum_{\gamma m} \left(
	\frac{ \bbra{\phi_{\alpha -n}} \hhat j_{0}^{u} \kket{\phi_{\gamma m}} \bbra{\phi_{\gamma m}} \hhat j_{0}^{v} \kket{\phi_{\alpha 0}}}{\omega + (\epsilon_\alpha -\epsilon_\gamma - m\Omega) + i0^+} 
	- \frac{ \bbra{\phi_{\alpha 0}} \hhat j_{0}^{v} \kket{\phi_{\gamma m}} \bbra{\phi_{\gamma m}} \hhat j_{0}^{u} \kket{\phi_{\alpha n}}}{\omega - (\epsilon_\alpha -\epsilon_\gamma - m\Omega) + i0^+} \right) 
	+ \mathbb{m}_{0\alpha\alpha}^{uv(n)} \bigg]&,
	\label{eq:sigman_diag}
\end{align}
where $u$ and $v$ are spatial directions. Using the Green's function defined in Eq.~(\ref{eq:FG}) one can also write
\begin{equation}\label{eq:sigman-FG}
    \bbsigma^{uv(n)}(\omega)=\frac{i}\omega\sum_\alpha g_{0\alpha} \bbra{\phi_{\alpha0}} \hhat{j}^u_0 \hhat{G}(\omega+\epsilon_\alpha+n\Omega) \hhat{j}^v_0
    +
    \hhat{j}^v_0 \hhat{G}^\dagger(-\omega+\epsilon_\alpha) \hhat{j}^u_0
    +
    \mathbb{m}^{uv}_0
    \kket{\phi_{\alpha n}}.
\end{equation}
\end{widetext}
Like the general susceptibility, Floquet optical conductivity satisfies the reality condition
\begin{equation}
    \bbsigma^{(-n)}(\omega) = \bbsigma^{(n)}(-\omega)^*.
\end{equation}

\subsection{Floquet optical sum rules}

The Floquet optical conductivity satisfies the general sum rule,
\begin{align}
    \frac1\pi\int_{-\infty}^\infty \bbsigma^{(n)}(\omega) d\omega = \left< \mathbb{m}^{(n)} \right>,
\end{align}
where the expectation value on the right hand side is the same as $\tr[g_0\mathbb{m}^{(n)}]$. Using the reality condition, we may also write this as
\begin{align}
    \frac1\pi\int_{0}^\infty [\bbsigma^{(n)}(\omega) + \bbsigma^{(-n)}(\omega)^*] d\omega = \left< \mathbb{m}^{(n)} \right>.
\end{align}
The sum rule can be obtained in the usual way~\cite{Shimizu_2011,Oshikawa_2019} from Eq.~(\ref{eq:oc}), $\bbsigma^{(n)}(\omega) = [\tilde\bbchi^{(n)}(\omega) - \left<\mathbb{m}^{(n)}\right>]/(i\omega)$, and the Kramers-Kronig relations~(\ref{eq:KK}) of susceptibility $\tilde\bbchi$. The integral over frequency is defined in the limit $\omega\to\omega + i0^+$, which maintains the poles of the susceptibility in complex frequency in the lower-half plane. Then, using $(\omega + i0^+)^{-1}=\mathcal{P}\omega^{-1} - i\pi\delta(\omega)$ and $\mathcal{P}\int_{-\infty}^\infty \frac{d\omega}\omega = 0$, we have
\begin{align}
    \frac1\pi\int_{-\infty}^\infty\bbsigma^{(n)}(\omega)d\omega &= 
    \frac1{i\pi} \mathcal{P}\int_{-\infty}^\infty \frac{\tilde\bbchi^{(n)}(\omega)-\left<\mathbb{m}^{(n)}\right>}{\omega} d\omega 
    \nonumber \\
    &~~~~- \left[\tilde\bbchi^{(n)}(0)-\left<\mathbb{m}^{(n)}\right>\right] \\ &=
    \left<\mathbb{m}^{(n)}\right>.
\end{align}

\subsection{Relation to Berry flux and Chern number of Floquet bands}
Eq.~(\ref{eq:sigman_diag}) can be understood as the spectral amplitude of excitations from side band $-n$ to the central FZ and from the central FZ to side band $n$ via virtual states $\ket{\phi_\gamma}$ in side band $m$. This is quite similar to the form of optical conductivity for a time-independent unperturbed Hamiltonian, except that one now needs to take account of FZ side bands. As in the time-independent case, we may seek a relationship between the DC Hall conductivity and the Chern number of the bands. In particular, setting $n=0$ 
\extra{we have
\begin{align}
\bbsigma^{uv(0)}(\omega;\tau_0) 
	&= \frac i{\omega} \sum_{\alpha}g_{0\alpha}  \left[ \sum_{\gamma m} \left(
	\frac{ j_{0\alpha\gamma}^{u(-m)}  j_{0\gamma\alpha}^{v(m)}}{\omega + (\epsilon_\alpha + m\Omega -\epsilon_\gamma) + i\eta} - \frac{ j_{0\alpha\gamma}^{v(-m)} j_{0\gamma\alpha}^{u(m)}}{\omega - (\epsilon_\alpha + m\Omega -\epsilon_\gamma) + i\eta} \right) \right] \\
	&= \frac i{\omega} \sum_{\alpha}g_{0\alpha}  \left[ \sum_{\gamma m} \left(
	\frac{ \bbra{\phi_{\alpha 0}} \hhat j_{0}^{u} \kket{\phi_{\gamma m}} \bbra{\phi_{\gamma m}} \hhat j_{0}^{v} \kket{\phi_{\alpha 0}}}{\omega + (\epsilon_\alpha -\epsilon_\gamma - m\Omega) + i\eta} 
	- \frac{ \bbra{\phi_{\alpha 0}} \hhat j_{0}^{v} \kket{\phi_{\gamma m}} \bbra{\phi_{\gamma m}} \hhat j_{0}^{u} \kket{\phi_{\alpha 0}}}{\omega - (\epsilon_\alpha -\epsilon_\gamma - m\Omega) + i\eta} \right) \right]
	\label{eq:sigman_diag_2} \\
	&= \frac i{\omega} \sum_{\alpha\gamma m} (g_{0\alpha} - g_{0\gamma}) 
	\frac{ \bbra{\phi_{\alpha 0}} \hhat j_{0}^{u} \kket{\phi_{\gamma m}} \bbra{\phi_{\gamma m}} \hhat j_{0}^{v} \kket{\phi_{\alpha 0}}}{\omega + (\epsilon_\alpha -\epsilon_\gamma - m\Omega) + i\eta},
\end{align}
Now,}
and taking the DC limit $\omega\to 0$, we find
\begin{align}
\bbsigma^{xy(0)}(0)
    = &-i \sum_{\alpha\neq \gamma,m}g_{0\alpha} 
	\frac{ \bbra{\phi_{\alpha 0}} \hhat j_{0}^{x} \kket{\phi_{\gamma m}} \bbra{\phi_{\gamma m}} \hhat j_{0}^{y} \kket{\phi_{\alpha 0}}
	}{(\epsilon_\alpha -\epsilon_\gamma - m\Omega)^2} \nonumber\\
	&-
	\{ x \leftrightarrow y \}.
\end{align}
The divergent $1/\omega$ terms vanish in the DC limit for the off-diagonal Hall conductivity (see Appendix for a proof). When the Floquet bands are either fully occupied or empty, this is the TKNN formula~\cite{ThoKohNig82a} that relates the DC Hall conductivity to the Chern number of the occupied bands $\kket{\phi_{\alpha0}}$, \emph{if} $\hhat{\vex j}$ is the current associated with the Bloch Hamiltonian of these bands, i.e. $\hhat H-\hhat Z_t$. This is in fact the case: $\hhat{\vex j} = \partial \hhat H/\partial \vex k = \partial (\hhat H - \hhat Z_t)/\partial \vex k$ since $\hhat Z_t$ is independent of $\vex k$.

To gain a better understanding of this quantization, we note that for any parameter $s$,
\begin{align}
    \bra{\phi_{\alpha}(t)}[\partial_{s} H(t)] \ket{\phi_{\gamma}(t)} 
    &= (\epsilon_\gamma - \epsilon_\alpha + i\partial_t) \inner{\phi_{\alpha}(t)}{\partial_{s} \phi_{\gamma}(t)} \nonumber \\
    &~~~+
    (\partial_s \epsilon_\alpha) \delta_{\alpha\gamma}.
\end{align}
Therefore, we can express the matrix elements of the current operator
\begin{align}
    \bbra{\phi_{\alpha0}}\hhat{j}_0^u\kket{\phi_{\gamma m}} &= \frac1T \int_0^T e^{im\Omega t} \bra{\phi_{\alpha}(t)}\partial_{k_u} H(t) \ket{\phi_{\gamma}(t)} dt \nonumber \\
    &\hspace{-10mm}= \frac{\epsilon_\gamma + m\Omega - \epsilon_\alpha}T \int_0^T e^{im\Omega t} \inner{\phi_{\alpha}(t)}{\partial_{k_u} \phi_{\gamma}(t)} dt \nonumber\\
    &\hspace{-10mm} \equiv -i (\epsilon_\gamma + m\Omega - \epsilon_\alpha) [r^u_{\alpha\gamma}]^{(m)}, \label{eq:jmr}
\end{align}
where we have defined the Fourier components of the elements of the ``position'' operator $r^u \equiv i \partial_{k_u}$, which furnishes the time-dependent Berry connection for the Floquet bands. Then,
\begin{align}\label{eq:DCHall0-rr}
\bbsigma^{xy(0)}(0) &= -i \sum_{\gamma\neq\alpha, m} g_{0\alpha} \left[ r^{x(m)}_{\alpha\gamma} r^{y(-m)}_{\gamma\alpha} - r^{y(m)}_{\alpha\gamma} r^{x(-m)}_{\gamma\alpha} \right] \\ &=
\sum_\alpha g_{0\alpha} F_\alpha^{(0)}, \label{eq:DCHall0}
\end{align}
where the Berry flux
\begin{align}
    F_{\alpha}(\vex k, t) =
     \partial_{k_x}r^y_{\alpha\alpha}(\vex k, t) - \partial_{k_y}r^x_{\alpha\alpha}(\vex k, t). \label{eq:FF}
\end{align}
Here, we are showing the dependence on the momenta explicitly for clarity, so $\alpha$ labels the bands and not their momenta. When the Floquet bands are fully occupied or empty and $g_{0\alpha}$ is independent of $\vex k$, we find $\bbsigma^{xy(0)} =\frac1{2\pi}\sum_{\alpha} g_{0\alpha} C_\alpha$
 with the Chern number $C_\alpha = \frac1{2\pi}\int F_\alpha^{(0)}(\vex k)d
\vex k$. Since the Floquet states at different times are unitarily related by the micromotion and Floquet spectrum is constant and gapped at all times, The Chern number $C_\alpha(t) = \frac1{2\pi}\int F_\alpha(\vex k,t)d\vex k = C_\alpha$ is time-independent.

We also show in the Appendix that the DC limit of Hall conductivity for Fourier modes $n\neq0$ is given by
\begin{widetext}
\begin{align}
    \bbsigma^{xy(n)}(0) &= 
    {-i} \sum_{\alpha\neq \gamma,m}g_{0\alpha} 
	\frac{ \bbra{\phi_{\alpha -n}} \hhat j_{0}^{x} \kket{\phi_{\gamma m}} \bbra{\phi_{\gamma m}} \hhat j_{0}^{y} \kket{\phi_{\alpha 0}}	-
	\bbra{\phi_{\alpha 0}} \hhat j_{0}^{v} \kket{\phi_{\gamma m}} \bbra{\phi_{\gamma m}} \hhat j_{0}^{u} \kket{\phi_{\alpha n}}
	}{(\epsilon_\alpha -\epsilon_\gamma - m\Omega)^2}  \\  &= 
    in\Omega\sum_{\gamma \alpha m} g_{0\alpha}\frac{
    r^{x(n+m)}_{\alpha\gamma}r^{y(-m)}_{\gamma\alpha}
    +
    r^{y(m)}_{\alpha\gamma}r^{x(n-m)}_{\gamma\alpha}
    }{\epsilon_\alpha-\epsilon_\gamma-m\Omega},\label{eq:DCHalln-rr}
\end{align}
and that, in the extended Floquet-Hilbert space, this expression can be recast in terms of the Floquet Green's function in the form,
\begin{equation}\label{eq:DCHalln-G}
    \bbsigma^{xy(n\neq0)}(0) = in\Omega\sum_{\alpha} g_{0\alpha} \bbra{\phi_{\alpha 0}}\partial_{k_x}\partial_{k_y}\hhat{G}^+_{\alpha n}
    +
    \{\partial_{k_x} \hhat G^-_{\alpha n} , \partial_{k_y}\}
    \kket{\phi_{\alpha n}},
\end{equation}
\end{widetext}
with $\hhat G^\pm_{\alpha n} = \frac12[\hhat G(\epsilon_\alpha + n\Omega) \pm \hhat G(\epsilon_\alpha)]$.
We note here that a finite value of this DC Hall conductivity signifies the heterodyne response of the Floquet system, i.e. the presence of a current at frequency $n\Omega$ in response to a DC electric field. We will show in a specific model that this response can be nonzero and large.

We should note here that the ideal filling of fully occupied or empty Floquet bands (i.e. $g_{0\alpha}=0$ or 1) is not necessarily obtained in experiments. This depends on the initial conditions at $t\to-\infty$ and on relaxation mechanisms that are not the subject of our study~\cite{Seetharam_2015,Dehghani1,Gavensky_2018}.  Moreover, the DC limit of the optical conductivity is not equivalent to a measurement of the DC conductivity with leads as the equilibrium states in the leads determine the filling of the Floquet states and reduce the DC conductivity from its quantized value~\cite{Farrell,Farrell2,Rodriguez-Vega2}.

\vspace{-5mm}
\section{Analytical Results}\label{sec:analytic}

In this section, we will first study a general driven two-band model, with the Hamiltonian
\newpage
\begin{equation}
    H_0(\vex k,t) = \vex d(\vex k) \cdot \gvex\sigma + \vex V(\vex k)\cdot\gvex\sigma \cos(\Omega t),
\end{equation}
where $\vex d(\vex k)$ specifies the static model with energy bands $\pm|\vex d(\vex k)|$ and
$\vex V(\vex k)$ is the amplitude of the external drive. In the following, we will suppress the explicit dependence on $\vex k$ for brevity. We obtain analytical expressions for Floquet Hamiltonian, micromotion, and current elements in a high-frequency approximation~\cite{EckAni15a,Rodriguez-Vega3}. We will use these expressions to calculate the Floquet optical conductivity in detail for specific models of driven systems.

\vspace{-5mm}
\subsection{Off-Resonant High-Frequency Approximation}\label{sec:offresHF}

Since the drive term commutes with itself at different times, we map the Hamiltonian to the rotating frame given by $U_R(t) = \exp[-i \vex V\cdot \gvex\sigma \sin(\Omega t)/\Omega]$. In this frame, the Hamiltonian is
\begin{align}
    H_R(t) &= U_R^\dagger(t)H(t)U_R(t) - iU_R^\dagger(t)\partial_t U_R(t) \nonumber\\
    &\equiv \vex d_R(t) \cdot\sigma,
\end{align}
with
\begin{align}
    \vex d_R(t) 
    &= \vex d_\parallel + \cos\left( \frac{2|\vex V|}\Omega \sin(\Omega t) \right) \vex d_\perp \nonumber\\
    &~~~+ \sin\left( \frac{2|\vex V|}\Omega \sin(\Omega t) \right) \vex d_\perp\times\hat{\vex V}.
\end{align}
Here, $\vex d_\parallel = (\vex d\cdot\hat{\vex V})\hat{\vex V}$ and $\vex d_\perp = \vex d - \vex d_\parallel$ are parallel and perpendicular components of $\vex d$ to $\vex V$ and the unit vector $\hat{\vex V} = \vex V/|\vex V|$.

Up to $1/\Omega$ in the high-frequency limit we find the Floquet Hamiltonian
\begin{align}
    H_F 
    &= H_R^{(0)}+\sum_{n\neq0}\frac{[H_R^{(-n)},H_R^{(n)}]+[H_R^{(0)},H_R^{(n)}-H_R^{(-n)}]}{2n\Omega} \nonumber\\
    &\equiv \vex d_F\cdot \gvex\sigma,
\end{align}
with
\begin{align}
    \vex d_F 
    &= \left[J_0(2|\vex V|/\Omega) - \frac{\pi \vex d\cdot\hat{\vex V}}{2\Omega} \mathrm{H}_0(2|\vex V|/\Omega) \right] \vex d_\perp \nonumber\\
    &+
    \left[\vex d\cdot\hat{\vex V} + \frac{\pi|\vex d_\perp|^2}{2\Omega}
    J_0(2|\vex V|/\Omega) \mathrm{H}_0(2|\vex V|/\Omega)
    \right] \hat{\vex V},
    \label{eq:offresdF}
\end{align}
where $J_0(z)$ and $\mathrm{H}_0(z)$ are Bessel and Struve functions. The micromotion in the original frame is $\Phi(t) = U_R(t)\Phi_R(t)$, where $\Phi_R(t) = \exp\left[\sum_{n\neq0}\frac{e^{in\Omega t}-1}{n\Omega} H_R^{(-n)} \right]$ is the micromotion in the rotating frame. However, since $H_R^{(n)}\propto (|\vex d_\perp|/\Omega)J_n(2|\vex V|/\Omega) \lesssim O(1/\Omega^2)$ for $|n|>0$, we have up to $1/\Omega$,  $\Phi(t) \approx U_R(t)$.

Using these expressions, the elements of the current operator in the spatial direction $u$ can be written as
$j^{Fu}_{\alpha\gamma}(t) = \sum_{m\in\mathbb{Z}}\bbra{\phi_{\alpha0}}\hhat{j}_0^u\kket{\phi_{\gamma m}}e^{-im\Omega t} = \vex j^{Fu}(t)\cdot
    \bra{\phi_\alpha}\gvex\sigma\ket{\phi_\gamma}$, with
\begin{align}
    \vex j^{Fu}(t)
    = 
    \vex j^u_\parallel(t) &+ \cos\left( \frac{2|\vex V|}\Omega \sin(\Omega t) \right) \vex j^u_\perp(t) \nonumber\\
    &+
    \sin\left( \frac{2|\vex V|}\Omega \sin(\Omega t) \right)
    \vex j^u_\perp(t) \times \hat{\vex V},\label{eq:offresJ}
\end{align}
where $\ket{\phi_\alpha}$ are the eigenstates of $H_F$, the current operator in the original frame is $\vex j^u(t) = \partial_{k_u}[\vex d + \vex V\cos(\Omega t)]$, and $\vex j^u_\parallel(t) = (\vex j^u(t)\cdot\hat{\vex V})\hat{\vex V}$ and $\vex j^u_\perp(t) = \vex j^u(t)-\vex j^u_\parallel(t)$ are its parallel and perpendicular components to $\vex V$.

\subsection{Resonant High-Frequency Approximation}\label{sec:resHF}

Now, we assume the frequency is small enough to satisfy the condition for resonance, $\Omega/2 = |\vex d|$. We will assume that the frequency is still large enough so that after, say, a single shift into the first Floquet zone, energy scales are small compared to the drive frequency. 

To obtain analytical expressions valid at and near resonance, we employ a resonant high-frequency approximation that accounts for resonant transitions. First, we switch to the rotating frame given by $U_R(t) = P_+ + P_-e^{-i\Omega t} = \exp[i\Omega t\: P_-]$, where $P_\pm = \frac12[\id \pm \hat{\vex d}\cdot\gvex\sigma]$ are the projectors to the two bands of the static model with energies $\pm|\vex d|$. The Hamiltonian in this rotating frame is
\begin{align}
    H_R(t) &= U_R^\dagger(t) H(t) U_R(t) - i U^\dagger(t) \partial_t U_R(t) \nonumber\\
    &\equiv \frac\Omega2 + \vex d_R(t)\cdot\gvex\sigma,
\end{align}
with
\begin{align}
    \vex d_R(t) &= \left( 1 - \frac\Omega{2|\vex d|} \right) \vex d + \frac12 \vex V_\perp + \vex V_\parallel \cos(\Omega t) \nonumber\\
    &~~~+ \frac12 \vex V_\perp \cos(2\Omega t) + \frac12 \vex V_\perp\times\hat{\vex d} \sin(2\Omega t).
\end{align}
Here, $\vex V_\parallel = (\vex V\cdot \hat{\vex d})\hat{\vex d}$ and $\vex V_\perp = \vex V - \vex V_\parallel$ are the parallel and perpendicular components of $\vex V$ to $\vex d$. Now, we obtain the high-frequency Floquet Hamiltonian,
\begin{align}
    H_F &= H_R^{(0)}+\sum_{n\neq0}\frac{[H_R^{(-n)},H_R^{(n)}]+[H_R^{(0)},H_R^{(n)}-H_R^{(-n)}]}{2n\Omega} \nonumber\\
    &\equiv \frac\Omega2 + \vex d_F\cdot \gvex\sigma,
\end{align}
with
\begin{align}\label{eq:resdF}
    \vex d_F = \left( 1 - \frac\Omega{2|\vex d|} + \frac{|\vex V_\perp|^2}{8\Omega|\vex d|} \right) \vex d + \frac12\left(\frac32 -\frac{|\vex d|}{\Omega}\right)\vex V_\perp.
\end{align}
It is worth noting here that the term $\frac34\vex V_\perp$ consists of a $\frac12\vex V_\perp$ contribution from $H_R^{(0)}$ \emph{and} a $\frac14\vex V_\perp$ contribution from commutators $[H^{(0)},H^{(\pm 2)}]/2\Omega$, which, nominally, belong to the next order in the $1/\Omega$ expansion. However, due to the resonant shift accounted for in the rotating frame, the $H_R^{(0)}$ component contains terms $\sim \Omega\hat{\vex d}$, which feeds back to the lowest order. Altogther, this yields the quasienergy bands $\frac\Omega2\pm|\vex d_F|$, 
with a gap at resonance  $\frac12|\vex V_\perp|\sqrt{1+(|\vex V_\perp|/4\Omega)^2}$~\cite{Wang,Farrell3}.

The micromotion operator in the original basis is given by $\Phi(t) = U_R(t)\Phi_R(t)$, where, a lengthy but straighforward calculation in the rotating frame, up to the same $1/\Omega^2$ order in the high-frequency expansion, yields~\cite{EckAni15a,Rodriguez-Vega3}
$
\Phi_R(t) = \exp\left[\sum_{n\neq0}\frac{e^{in\Omega t}-1}{n\Omega} H_R^{(-n)} \right] \equiv \exp[i\gvex\alpha(t)\cdot \gvex\sigma],
$
with
\begin{align}
    \gvex\alpha(t) 
    &= \frac{\sin(\Omega t)}{2\Omega}\left[2\vex V_\parallel + \cos(\Omega t) \vex V_\perp + \sin(\Omega t) \vex V_\perp\times \hat{\vex d} \right] \nonumber\\
    &\equiv \frac{\tilde V\sin(\Omega t)}{2\Omega}\hat{\gvex\alpha}(t).
    \label{eq:RWmicro}
\end{align}
Here, $\hat{\gvex\alpha}(t) =  \left[2\vex V_\parallel + \cos(\Omega t) \vex V_\perp + \sin(\Omega t) \vex V_\perp\times \hat{\vex d} \right]/\tilde V$ is a unit vector.
We note that the vector form of $\gvex\alpha(t)$, shown in the square brackets in Eq.~(\ref{eq:RWmicro}), consists of a component of fixed magnitude $|\vex V_\perp|$ rotating perpendicular to $\vex d$, and a fixed component $2\vex V_\parallel$ parallel to $\vex d$. Thus, the magnitude of this vector, $\tilde V = \sqrt{4|\vex V_\parallel|^2+|\vex V_\perp|^2}$, is constant in time.

Using these expressions, we find the elements of the current operator in spatial direction $u$ in the Floquet basis as
$j^{Fu}_{\alpha\gamma}(t) = \sum_{m\in\mathbb{Z}}\bbra{\phi_{\alpha0}}\hhat{j}_0^u\kket{\phi_{\gamma m}}e^{-im\Omega t} = \vex j^{Fu}(t)\cdot\bra{\phi_\alpha}\gvex\sigma\ket{\phi_\gamma}$ with
\begin{align}
    \vex j^{Fu}(t)
    =
    \vex j_{R\parallel}^u(t)
    &+
    \cos\left(\frac{\tilde V}\Omega \sin(\Omega t)\right)
    \vex j^u_{R\perp}(t) \nonumber\\
    &+
    \sin\left(\frac{\tilde V}\Omega\sin \Omega t\right) 
    \hat{\gvex\alpha}(t) \times
    \vex j^u_{R\perp}(t),
    \label{eq:resJ}
\end{align}
where $\vex j^u_{R\parallel}(t)=[\vex j^u_{R}(t)\cdot\hat{\gvex\alpha}(t)]\hat{\gvex\alpha}(t)$ and $\vex j^u_{R\perp}(t)=\vex j^u_{R}(t)-\vex j^u_{R\parallel}(t)$ are parallel and perpendicular components to $\gvex\alpha$ of the current operator $\vex j^u_R\cdot\gvex\sigma$ in the spatial direction $u$ in the rotating frame,
\begin{equation}
    \vex j^u_R(t) = \vex j^u_\parallel(t) + \cos(\Omega t) \vex j^u_\perp(t) + \sin(\Omega t) \hat{\vex d}\times\vex j^u(t),
\end{equation}
and $\vex j^u_\parallel(t) = [\vex j^u(t) \cdot \hat{\vex d}]\hat{\vex d}$ and $\vex j^u_\perp(t) = \vex j^u(t) - \vex j^u_\parallel(t)$ are, in turn, parallel and perpendicular components to $\vex d$ of the current operator $\vex j^u(t)\cdot\gvex\sigma$ in the original frame with $\vex j^u(t) = \partial_{k_u}[\vex d + \vex V \cos(\Omega t)].$ 
Here, the Floquet modes $\ket{\phi_\alpha}$ are eigenstates of the Floquet Hamiltonian~(\ref{eq:resdF}).
When $\vex V$ is independent of $\vex k$, the current operator in the original frame simplifies to $\vex j^u=\partial_{k_u}\vex d$, which is time-independent. 

\subsection{Dirac Electrons Irradiated with Linearly Polarized Light}

\begin{figure}[t]
   \centering
   \includegraphics[width=2.2in]{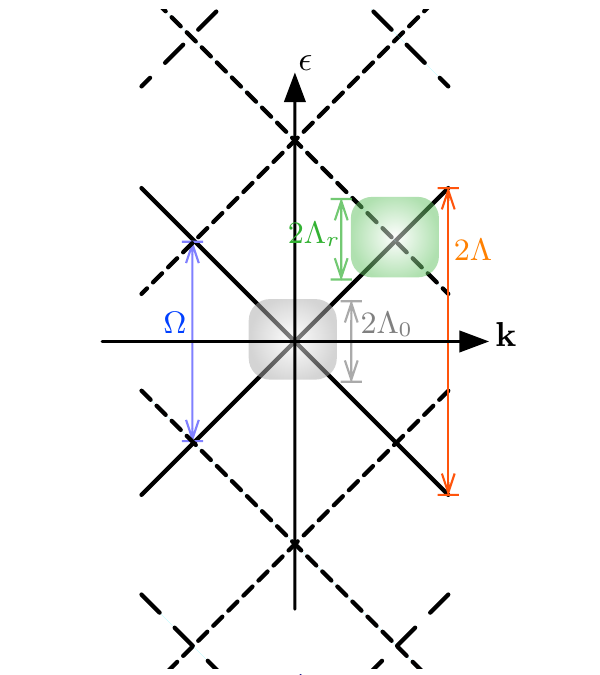}
   \caption{Extended Floquet zones and cut-offs for resonantly driven Dirac cone. The cut-off $\Omega/2<\Lambda<\Omega$ isolates a single resonance in the spectrum. The cut-offs $\Lambda_0\ll \Omega/2$ at the original Dirac point and $\Lambda_r\ll\Omega/2$ at the resonant Dirac point select the dominant contributions to the optical response in each case.}
   \label{fig:DLcut}
\end{figure}

As our first example, we choose $\vex d = v(k_x,k_y,0)$ and $\vex V = (V,0,0)$. This can be taken to represent Dirac electrons with Fermi velocity $v$ driven by a linearly-polarized laser field, which is realized in irradiated graphene or the surface of a strong topological insulator. 
First, we note that along $k_y=0$, we may find the full evolution operator $U(t) = e^{-i [vk_x t-(V/\Omega)\sin(\Omega t)]\sigma_x}$, which yields the Floquet Hamiltonian $H_F = vk_x\sigma_x$. Thus, the original Dirac point remains gapless and there is also a pair of gapless points at resonance for $\vex k = \vex k_r^{\pm }=(\pm\Omega/2v,0)$ at any frequency. We will see below that analyzing this system more generally requires both off-resonant and resonant treatments of the Floquet states. For simplicity, we set the bandwidth by choosing an energy cut-off $\Lambda$ such that $\Omega/2 < \Lambda < \Omega$, which ensures there is a single resonance at $\vex k = \vex k_r$ with $v|\vex k_r|=\Omega/2$. The extended Floquet zones and the relevant cut-offs are sketched in Fig.~\ref{fig:DLcut}.

Using Eq.~(\ref{eq:offresdF}) for off-resonant high-frequency approximation, we find that the original Dirac point  becomes slightly anisotropic with Floquet Hamiltonian $H_F \approx vk_x\sigma_x + vJ_0(2V/\Omega)k_y\sigma_y$.
To obtain the Floquet spectrum near resonance, we use the resonant high-frequency approximation and Eq.~(\ref{eq:resdF}), noting $\vex V_\parallel = Vk_x(k_x,k_y,0)/|\vex k|^2$ and $\vex V_\perp = V k_y (k_y,-k_x,0)/|\vex k|^2$. 
We can see that the gap at resonance has the magnitude $(vk_{ry}  V/\Omega)\sqrt{1+(v k_{ry}  V/2\Omega^2)^2}$. This gap is maximized for $\vex  k = (0,\pm\Omega/2v)$ at a value $(V/2)\sqrt{1+(V/4\Omega)^2}$ and closes at $\vex k_r^{\pm}$, as it must. Expanding around $\vex k = \vex k_r^{\pm}+\vex q$ for small $|\vex q|\ll \Omega/v$, we find two highly anisotropic Dirac points with 
the Floquet Hamiltonian $H_F \approx \Omega/2 + v q_x \sigma_x \mp v(V/\Omega) q_y \sigma_y$.

The current operators in the original frame are $\vex j^x = v(1,0,0)$ and $\vex j^y = v(0,1,0)$. The Floquet current elements can be written as $j^{Fu}_{\alpha\beta} = \vex j^{F u}\cdot\bra{\phi_\alpha}\gvex\sigma\ket{\phi_\beta}$, where $\vex j^{Fu}$ are found from Eqs.~(\ref{eq:offresJ}) or~(\ref{eq:resJ}). At the original Dirac point, we find $\vex j^{Fx}=v(1,0,0)$ and $\vex j^{Fy} = v\cos[(2V/\Omega)\sin(\Omega t)](0,1,0)-v\sin[(2V/\Omega)\sin(\Omega t)](0,0,1)$. The Fourier components are, thus,
\begin{align}
    \vex j^{Fx(n)} &= v\delta_{n0} (1,0,0)  \\
    \vex j^{Fy(n)} &= \begin{cases} v J_n(2V/\Omega)(0,1,0) & \text{even}~n, \\ 
    - iv J_n(2V/\Omega)(0,0,1) & \text{odd}~n.
    \end{cases}
\end{align}
To proceed analytically, we set an energy limit $v\sqrt{k_x^2+[J_0(2V/\Omega)k_y]^2}<\Lambda_0 < \Omega/2$.
After rescaling $J_0(2V/\Omega)k_y \mapsto k_y$, we find
\begin{equation}
    \sigma^{xx(n)}_0(\omega) = \frac{\delta_{n0}}{J_0(2V/\Omega)}\sigma_D,
\end{equation}
where $\sigma_D = 1/16$ is the optical conductivity of a single half-filled Dirac cone (in units of $e^2/\hbar$)\cite{Gusynin}. 
Assuming $\Lambda_0 \ll \Omega$, we see that the main contribution to optical conductivity in Eq.~(\ref{eq:sigman_diag}) is found when $\omega\pm m\Omega \approx \epsilon_{\alpha}-\epsilon_\gamma$ in the central Floquet zone. Thus, after some algebra, we find for $|\omega-m\Omega|\lesssim\frac\Omega2$,
\begin{equation}
    \sigma^{yy(n)}_0(\omega) \approx \zeta_{m}\frac{J_{n+m}(2V/\Omega)J_m(2V/\Omega)}{J_0(2V/\Omega)}\frac{\omega-m\Omega}{\omega}\sigma_D,
\end{equation}
with  $\zeta_{m} = [3-(-1)^{m}]/2$ for even $n$. For odd $n$, $\sigma^{yy(n)}_0 = 0$.  We note that the factor $\zeta_m=2$ for odd $m$ arises from the fact that $\bra{\vex k -}\sigma_z\ket{\vex k +} = 1 = |\bra{\vex k -}\sigma_x\ket{\vex k +}|^2 + |\bra{\vex k -}\sigma_y\ket{\vex k +}|^2$ where $\ket{\vex k \pm}$ are the eigenstates of $H_F$ with quasienergy $\pm v \sqrt{k_x^2+[J_0(2V/\Omega)k_y]^2}$. For small $V/\Omega\lesssim 1$, the typical values of this contribution scale as $\sigma^{yy(n)}_0\sim (V/\Omega)^{n+2m}/[(n+m)! m! m]$ and vanish quickly with increasing $n$ and $m$. Optical Hall conductivity $\sigma^{xy(n)}_0(\omega)=0$ since it involves integrals over odd functions of $\vex k$. This is similar to the situation for a static Dirac cone, as in graphene. 

\begin{figure}[t]
   \centering
   \includegraphics[width=3.45in]{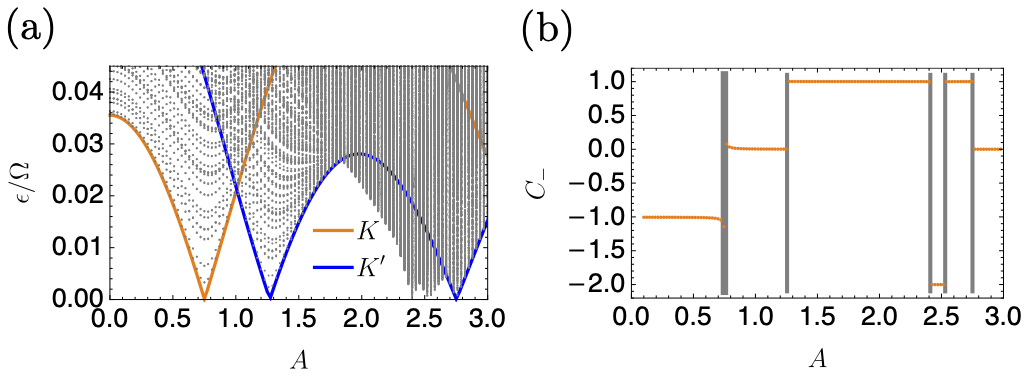}
   \caption{Driven Haldane model. Quasienergies close to Floquet zone center (a) and Chern number of the lower band (b) as a function the intensity $A$ for $\Omega/J = 9$. Changes in the Chern number are accompanied by gap closings at the Floquet zone center via the appearance of single or multiple Dirac points at high-symmetry or other points in the Brillouin zones.}
   \label{fig:Haldane_1}
\end{figure}

Keeping terms linear in $\vex q = \vex k - \vex k_r^{\pm}$ near the resonant Dirac points, it is easy to see that only current elements at $\vex q = 0$ make nonzero contribution to optical conductivity, since integrals over odd functions of $\vex q$ vanish. The current elements are
%
\begin{align}
    \vex j^{F x\pm}_r &= v(1,0,0), \\
    \vex j^{F y\pm}_r &= v \cos\left[(2V/\Omega)\sin(\Omega t)\pm\Omega t\right] (0,1,0) \nonumber\\
    &~~~+ \sin\left[(2V/\Omega)\sin(\Omega t)\pm\Omega t\right](0,0,1).
\end{align}
As with the original Dirac point, we rescale the anisotropic momentum $(V/\Omega) q_y \mapsto q_y$ to find
\begin{equation}
    \sigma^{xx(n)\pm}_r(\omega) = \delta_{n0}\frac{\Omega}{V}\sigma_D,
\end{equation}
and $\sigma^{xy(n)\pm}_r(\omega)=0$. Thus, the anisotropic Dirac point in the Floquet spectrum can be identified via a large contribution to the optical longitudinal conductivity in the direction parallel to the polarization. 

To find $\sigma^{yy(n)\pm}_r$, we proceed again by setting an energy cut-off $v\sqrt{q_x^2+[Vq_y/\Omega]^2}<\Lambda_r\ll \Omega/2$ and find, after some algebra, that while for odd $n$ contributions from the two resonant Dirac points have opposite signs and cancel, they add up for even $n$:
\begin{align}
    \sigma^{yy(n)\pm}_r(\omega) &\approx \frac{\Omega}{4V}\left[
    \zeta_m 
    J^-_m(2V/\Omega)
    J^-_{n+m}(2V/\Omega) \right.\nonumber\\
    &\left.
    -\zeta_{m+1}
    J^+_m(2V/\Omega)
    J^+_{n+m}(2V/\Omega)
    \right] \frac{\omega-m\Omega}{\omega}\sigma_D,
\end{align}
where $J^\pm_\nu(z) = J_{\nu-1}(z) \pm J_{\nu+1}(z)$ and 
$|\omega-m\Omega|\lesssim\frac\Omega2$. For small $V/\Omega\lesssim 1$, some typical values of this contribution are $\sigma^{yy(0)\pm}_r(\omega)\approx (v/\Omega)\sigma_D$ for $m=0$ and $\sigma^{yy(0)\pm}_r(\omega) \approx \frac{\omega-m\Omega}{\omega} (\Omega/4V) \sigma_D$ for $m=\pm1$, which show, respectively, suppression and enhancement
by the anisotropy of the resonant Dirac point. Interestingly, the second Fourier components   $\sigma^{yy(-2m)\pm}_r(\omega) \approx -3\sigma^{yy(0)\pm}_r(\omega)$ for $m=\pm1$ is even larger in magnitude, but higher Fourier components $|n|>2$ and larger $|m|>1$ are quickly suppressed.

\begin{figure}[t]
   \centering
   \includegraphics[width=3.45in]{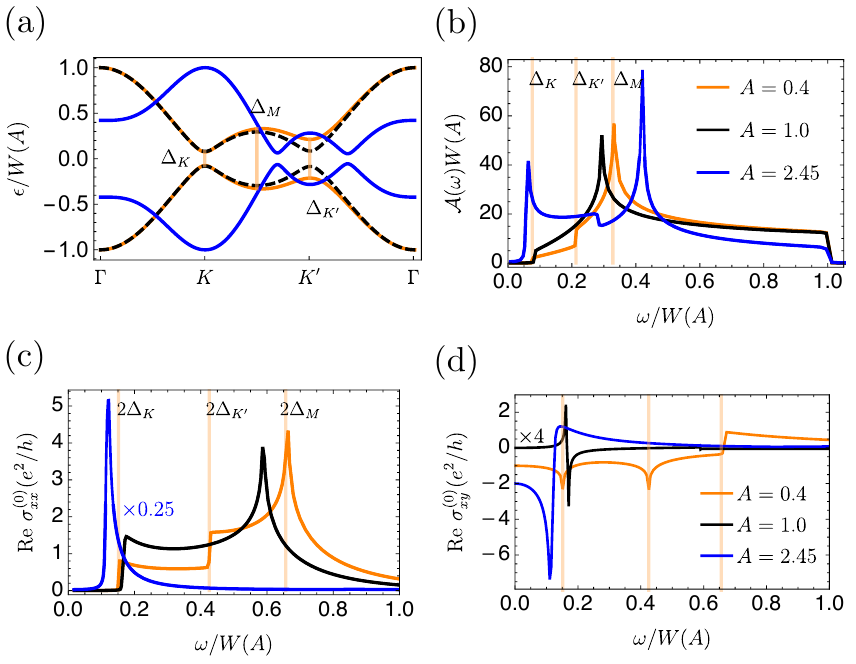}
   \caption{Spectral properties and homodyne optical conductivty of the driven Haldane model at $\Omega/J = 9$. (a) Quasienergies along a high-symmetry path in the BZ for three different laser intensities, $A=0.4$, $A=1$, and $A=2.45$. (b) Density of states $\mathcal{A}(\omega)$ as a function of probe frequency.
   (c) Longitudinal conductivity $\sigma^{(0)}_{xx}(\omega)$  and (d) Hall conductivity $\sigma^{(0)}_{xy}(\omega)$ as a function of probe frequency $\omega$ normalized to the quasienergy bandwidth for $\Omega/J = 9$. The DC Hall conductivity agrees with the Chern numbers in Fig.~\ref{fig:Haldane_1}(b). The half-bandwidths are $W(A=0.4)/J = 2.89$, $W(A=1)/J = 2.30$, and $W(A=2.45)/J = 0.52$. In panels (c) and (d), the scaling factor indicated in blue corresponds to the case $A=2.45$.  }
   \label{fig:Haldane_2}
\end{figure}

\begin{figure*}[t]
   \centering
   \includegraphics[width=5.5in]{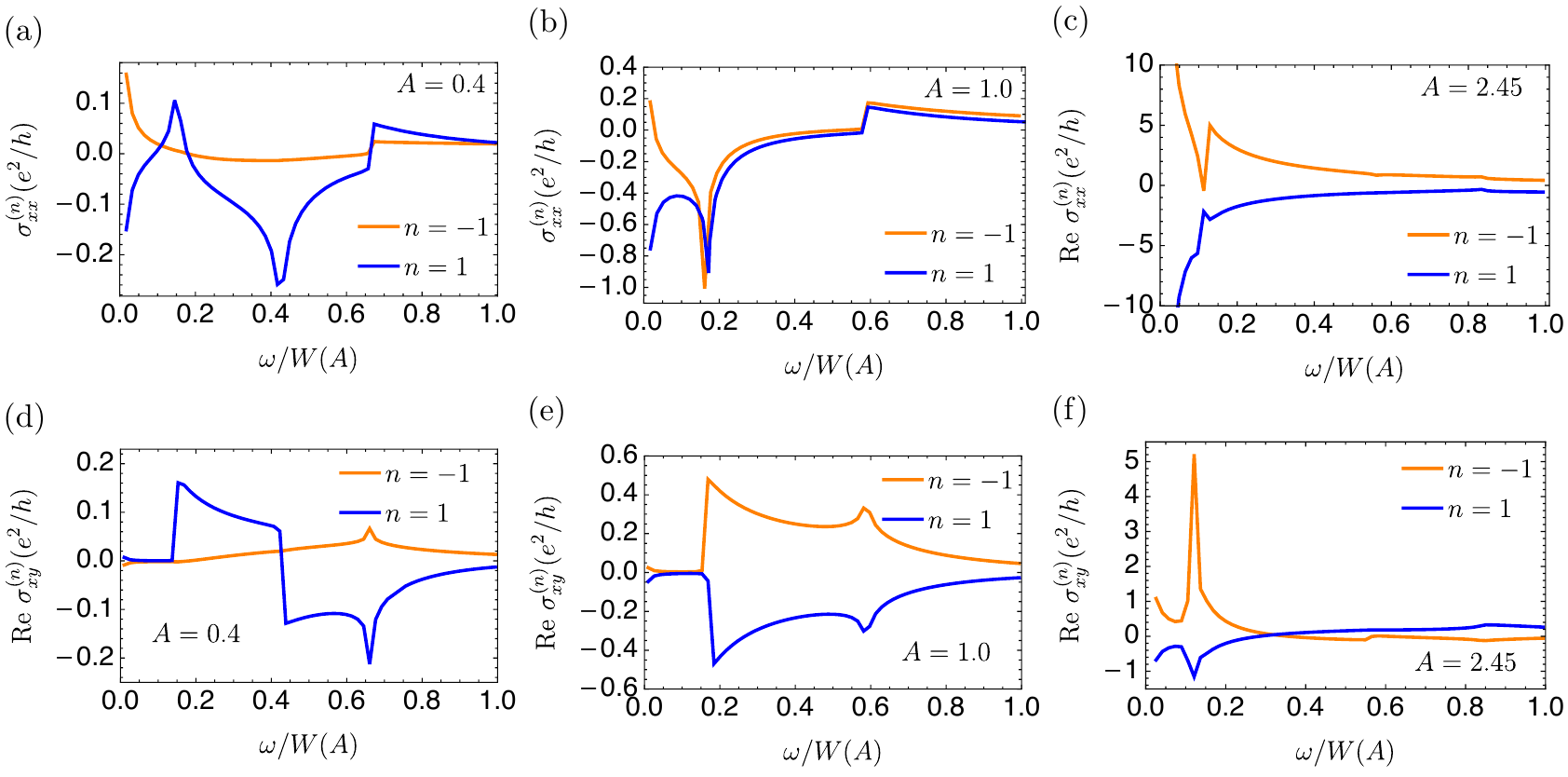}
  \caption{Heterodyne optical conductivity of the driven Haldane model at $\Omega/J=9$ and three different intensities of the laser. %
  }
   \label{fig:Haldane_4}
\end{figure*}

\section{Numerical Results}\label{sec:numerics}
In this section, we present our numerical results obtained by exact numerical solutions of the Floquet states as well as numerical integration of our analytical expressions obtained in the previous section in more complicated two-band models. We consider two examples: the Haldane model~\cite{Haldane} driven by circularly polarized light, and the driven Bernevig-Hughes-Zhang (BHZ) model of a driven quantum well~\cite{Bernevig}. 

\subsection{Driven Haldane Model}
As our next example, we consider the Haldane model~\cite{Haldane} driven by circularly-polarized light. Setting the nearest-neighbor distance on the honeycomb lattice $a=1$ and with periodic boundary conditions, the tight-binding Hamiltonian is given by
\begin{align}
H_0(\vex k,t) = J \sum_{j=1}^3 &\left\{ \cos \left[ \vex k(t)\cdot \vex a_j \right] \sigma_x - 
\sin \left[\vex k(t) \cdot \vex a_j \right] \sigma_y \right. \nonumber\\
&\left.+~\delta' \sin  \left[\vex k(t) \cdot \vex b_j \right] \sigma_z
\right\} + \mu_s \sigma_z ,
\end{align}
where $\vex k(t) = \vex k - \vex A (t)$ with the circularly polarized vector potential $\vex A(t) = A(\cos \Omega t, \sin \Omega t,0)$. $J$ is the hopping amplitude to the nearest neighbors ${\bf a}_j = \left( \cos \theta_j , \sin \theta_j \right)$ with $\theta_j = \pi/3(2j-1)$, and
$\delta'$ is the ratio of the hopping amplitude with $J$ to the next-nearest neighbors at  $\vex b_1 = \vex a_3 - \vex a_2$, $\vex b_2 = \vex a_2 - \vex a_1$, and  $\vex b_3 = \vex a_1 - \vex a_3$. Here, the Pauli matrices act on the sublattice space, and $\mu_s$ is the staggered chemical potential. We will denote half of the Floquet bandwidth by $W$, which is related to $J,\delta'$ and varies with the drive parameters $A,\Omega$. Note that for $
\delta' = 0$ and $\mu_s=0$ we obtain the tight-binding model for graphene. %
As is well known, in equilibrium ($A=0$) the Haldane  model exhibits non-trivial topological phases: for example, when $0<|\delta'| < 1/3$ and $|\mu_s/J| < 3 \sqrt{3} \delta'$, the bands have Chern numbers $C=\pm 1$.

We calculate the Floquet spectrum in the extended Hilbert space, 
as detailed in previous sections. The Fourier components of the Hamiltonian are given by
\begin{align}
&H_0^{(n)} = J \sum_{j=1}^3  \left[ \Upsilon^{(n)}_{j+} (A) \sigma_x - i \Upsilon^{(n)}_{j-} (A) \sigma_y - 2 i \delta' Y^{(n)}_{j-} (A) \sigma_z\right]
\nonumber \\ 
&\hspace{2in}+
\mu_s \delta_{n0} \sigma_z 
,
\end{align}
where 
\begin{align}
\Upsilon^{(n)}_{j\pm} (A) & = \frac{1}{2i^n} \left[e^{i\vex k \cdot \vex a_j} \pm (-1)^n
e^{-i\vex k \cdot \vex a_j}\right] e^{-i n \theta_{1-j}}  J_n(A),\\
Y^{(n)}_{j\pm} (A) & = \frac{1}{2} \left[e^{i\vex k \cdot \vex b_j} \pm (-1)^n
e^{-i \vex k \cdot \vex b_j}\right] e^{in \theta_j} J_n(\sqrt{3}A).
\end{align}

The Fourier components of the current operator matrix elements entering in Floquet optical conductivity are $\vex j_{0\alpha \beta}^{F(n)} = \sum_{lm} \langle  \phi^{(l)}_\alpha | \vex j_0^{(n+l-m)} | \phi^{(m)}_\beta \rangle $, where $\vex j_0^{(n)}$ is the Fourier component of $\vex j(t) = \partial H_0(t)/\partial \vex k$, given by
\begin{align}
\vex j_0^{(n)} = J \sum_{j=1}^3\left\{ \vex a_j \left[ i  \Upsilon^{(n)}_{j-} (A) \sigma_x +  \Upsilon^{(n)}_{j+} (A) \sigma_y \right]\right.& \nonumber\\
\left.+ 2 \delta' \vex b_j Y^{(n)}_{j+} (A) \sigma_z
\right\}&.
\end{align}

\begin{figure*}[t]
   \centering
   \includegraphics[width=7in]{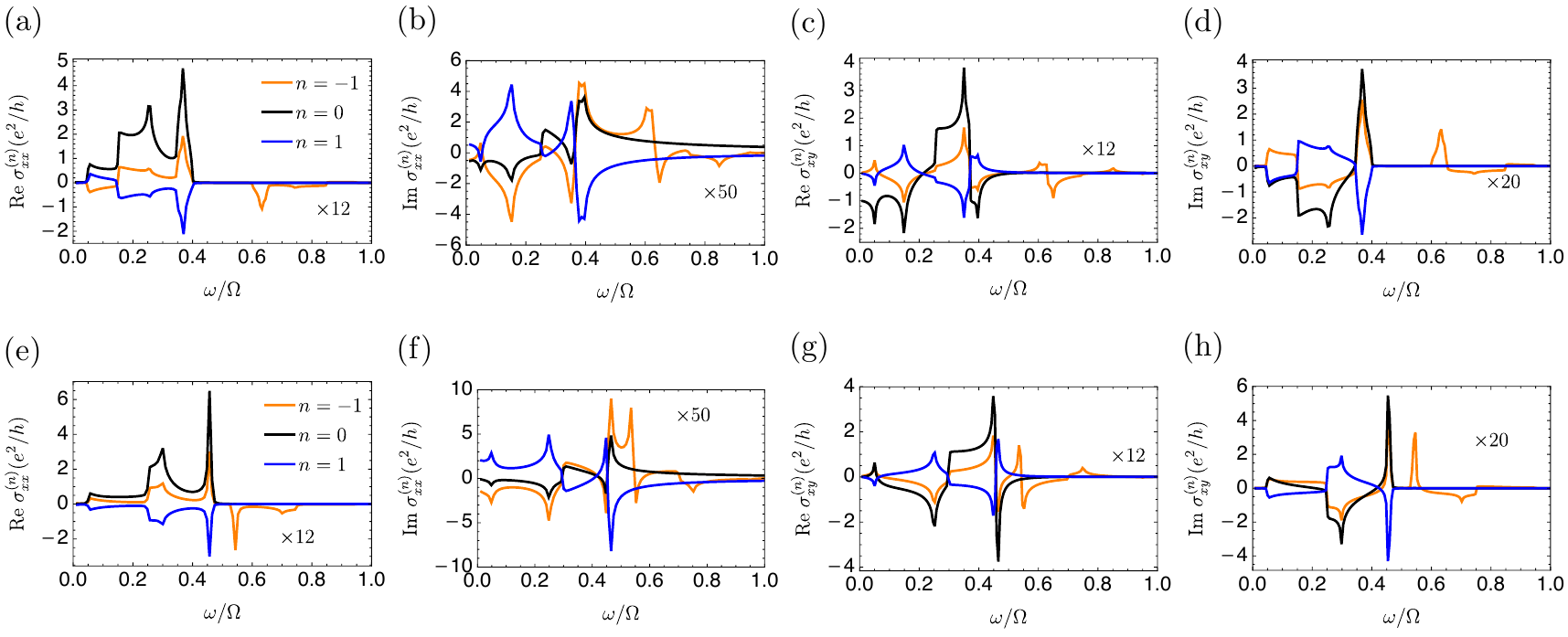}
   \caption{Homodyne and heterodyne optical conductivity of the driven quantum well model in the high-frequency limit, $\Omega/A=8$. The other parameters are chosen to be $B/A = 0.2$, $V/A=0.35$. In panels (a-d), $M/A=0.2$ corresponds to the topological phase of the equilibrium (average) Hamiltonian. In panels (e-h), $M/A=-0.2$ corresponds to the trivial phase of the equilibrium (average) Hamiltonian. In each panel, the scaling factor applies to $n=\pm1$. %
    }
   \label{fig:qw_1}
\end{figure*}
\begin{figure*}[t]
   \centering
   \includegraphics[width=7in]{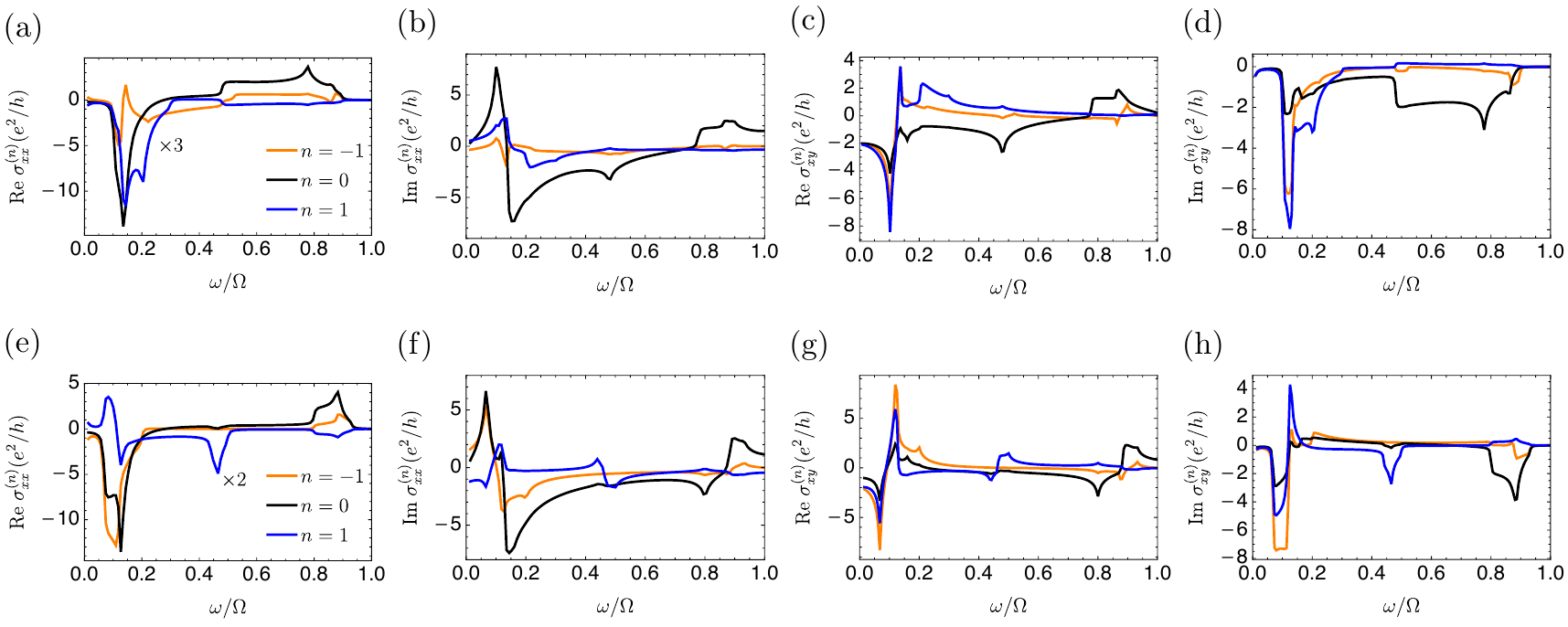}
    \caption{Homodyne and heterodyne optical conductivity of the driven quantum well model in the mid-frequency limit, $\Omega/A=2.5$. The other parameters are as in Fig.~\ref{fig:qw_1}.  In panels (a) and (e), the scaling factor applies to $n=\pm1$.%
    }
   \label{fig:qw_2}
\end{figure*}

For concreteness, in the following we will fix the values $\mu_s/J=0.2$ and $\delta' = 0.1 $, which place the system in equilibrium in the topological phase. For the driven system, we set $\Omega/J = 9$, which is larger than the bandwidth $2W(A=0)$, so for weak driving we expect the Floquet spectra to be approximately the same as in equilibrium. Fig.~(\ref{fig:Haldane_1}a) shows the quasienergies around the Floquet zone center $\epsilon/\Omega = 0$ as a function of amplitude $A$. For this drive frequency and the range of amplitudes shown, there are no gap closings at the Floquet zone edge $\epsilon=\Omega/2$. Due to inversion symmetry breaking caused by nonzero $\mu_s$, the gap at the $K$ and $K'$ points close at different values of drive amplitude: the gap at the $K$ point closes first at $A \approx 0.75$, followed by the gap at $K'$ near $A\approx1.3$.
Other gap closings occur near $A\approx 2.4$ and $A\approx2.5$, which involve three points consistent with the three-fold rotational symmetry of the honeycomb lattice.

In Fig.~(\ref{fig:Haldane_1}b) we plot, as a function of the drive amplitude $A$, the Chern number of the lower band,
$
C_- = \frac{1}{2 \pi} \int %
F_-(\vex k, t) d \vex k,
$
as defined in Eq.~(\ref{eq:FF}).
Changes in the Chern number are concomitant with gap closings. When the gap closes at a single $\vex k$, such as $K$ or $K'$ points, the Chern number changes by one caused by the passage through a Dirac cone in the gapless spectrum~\cite{Kundu2,Kundu3}. On the other hand, as mentioned above gap closings near $A \approx 2.5$ occur at three points,  inducing changes in the Chern number by 3. As representative case studies, we will consider three amplitudes below: $A = 0.4$ corresponding to $C_-=-1$ as in equilibrium, $A=1$ corresponding to $C_-=0$ following the Floquet gap closing at $K$ point, and $A=2.45$ corresponding to $C_-=-2$ in the lower Floquet band. %

In Fig. (\ref{fig:Haldane_2}a), we plot the quasienergy spectrum along a high-symmetry path in the Brillouin zone (BZ) for three laser intensities. In Fig. (\ref{fig:Haldane_2}b) we plot the average density of states in a drive cycle~\cite{Oka} 
$
\mathcal{A}(\omega) = -\frac1\pi \im \Tr[\hat G^{(0)}(\omega)] 
$,
where the Floquet Green's function is defined in Eq.~(\ref{eq:FGn}). Now, we calculate the optical conductivity tensor $\bbsigma^{(n)}(\omega)$ as defined in Eq.~(\ref{eq:sigman_diag}), assuming ideal occupation of the lower Floquet bands, i.e. $g_{0 \alpha \beta} = \delta_{\alpha \beta} \Theta(-\epsilon_\alpha)$. The results for the longitudinal and Hall conductivities are shown in Fig.~(\ref{fig:Haldane_2}c) and Fig.~(\ref{fig:Haldane_2}d) for $n=0$ and in Fig.~(\ref{fig:Haldane_4}) for $n=\pm1$.

The structure of peaks and steps in the response can be understood as arising from particle-hole excitations near van Hove singularities, which in our particle-hole symmetric spectrum is equal to twice the energies in the single-particle Floquet spectrum. The latter can be seen in the spectral density $\mathcal{A}(\omega)$. We note that for $A = 2.45$, $\mathcal{A}(\omega)$ shows a peak at $\omega\approx 0.42 W$, but the corresponding structure near $\omega \approx 0.84 W$ is not visible in optical conductivity since most of the optical weight for this amplitude is shifted close to the topological gap. As expected, the DC Hall conductivity is $\sigma^{(0)}_{xy}(0)= C_-$ (in units of $e^2/h$). 

In order to perform a more detailed study of the Fourier components of the conductivity tensor, and its real and imaginary components, in the next section we consider the driven quantum well, where we can employ our analytical expressions more easily.  

\begin{figure}[t]
   \centering
   \includegraphics[width=3.45in]{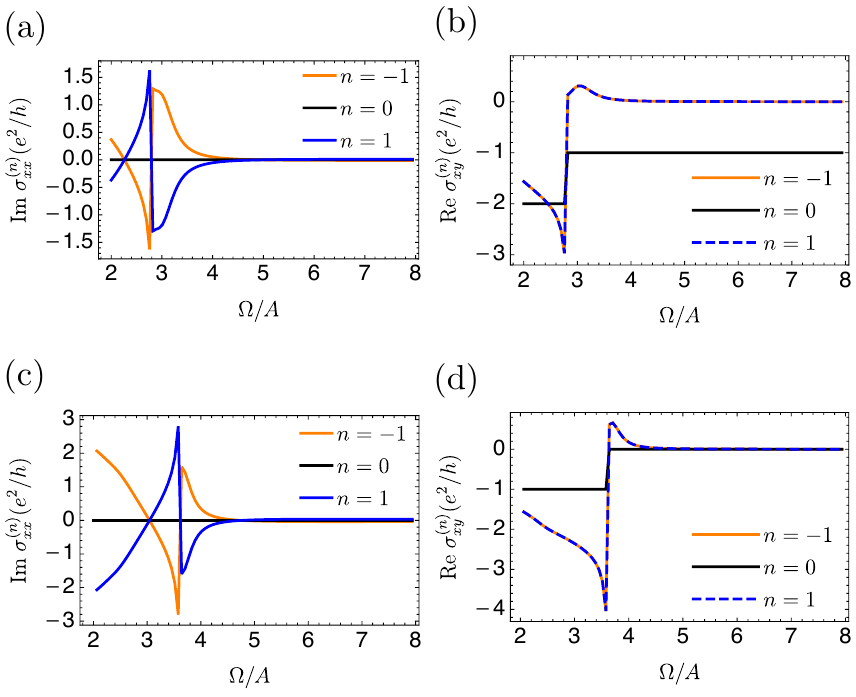}
    \caption{The DC heterodyne optical conductivity of the driven quantum well as a function of the drive frequency. In this case, $\re \sigma^{(n)}_{xx} =\im \sigma^{(n)}_{xy}=0$ in the range of drive frequencies considered. In panels (a) and (b) $M/A=0.2$, and in panels (c) and (d) $M/A=-0.2$. 
   The other parameters are as in Figs.~\ref{fig:qw_1} and~\ref{fig:qw_2}. 
   }
   \label{fig:heterodyne_3}
\end{figure}

\subsection{Driven Quantum Well}
As our final example, we take
$\vex d = (A \sin k_x, A\sin k_y, M-4B + 2B \cos k_x + 2B \cos k_y)$ and $V=(0,0,V)$ to represents a two-band model of a driven quantum well, such as one formed in a semiconductor heterojunction~\cite{Bernevig} and driven by an ac gate voltage. The equilibrium model for $V =0$ has a topological phase when $\sgn(MB)>0$, characterized by a nonzero Chern number $C = \frac1{4\pi} \int d\vex k\: \hat{\vex d} \cdot \partial_{k_x}\hat{\vex d} \times \partial_{k_y}\hat{\vex d} = \frac12[1+\sgn(MB)]$. For concreteness, we assume below $A>0$, $B/A=0.2$, $V/A=0.35$, and $M/A = \pm 0.2$ with $M<0$ corresponding to the trivial phase, $C=0$ and $M>0$ to the topological phase, $C=1$ in equilibrium.

In contrast to the driven Haldane model, our first example, the current operators are time-independent,
$
    j^u = A \cos k_u \sigma_u - 2 B \sin k_u \sigma_z,
$
for $u=x,y$. 
The diamagnetic response depends on the inverse-mass matrix
$\mathbb{m}_{0}^{uv} = -\delta_{uv}(A \sin k_u \sigma_u + 2B \cos k_u \sigma_z)$, which is diagonal in the spatial directions and, therefore, do not contribute to the Hall response.

\begin{figure}[t]
   \centering
   \includegraphics[width=3.45in]{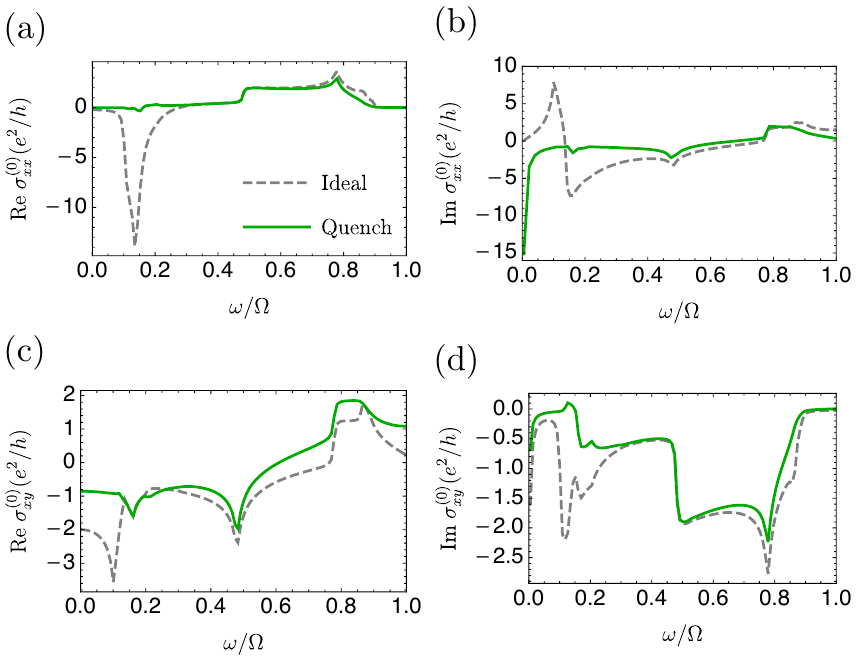}
   \caption{Homodyne optical conductivity of the driven quantum well as a function of the probe frequency in the steady state following a quench by the drive (solid curve) compared to the ideal Floquet ocupation (dashed curve). The parameters are as in Fig.~\ref{fig:qw_2}(a-d).}
   \label{fig:qw_3}
\end{figure}

\begin{figure*}[t]
   \centering
   \includegraphics[width=6.9in]{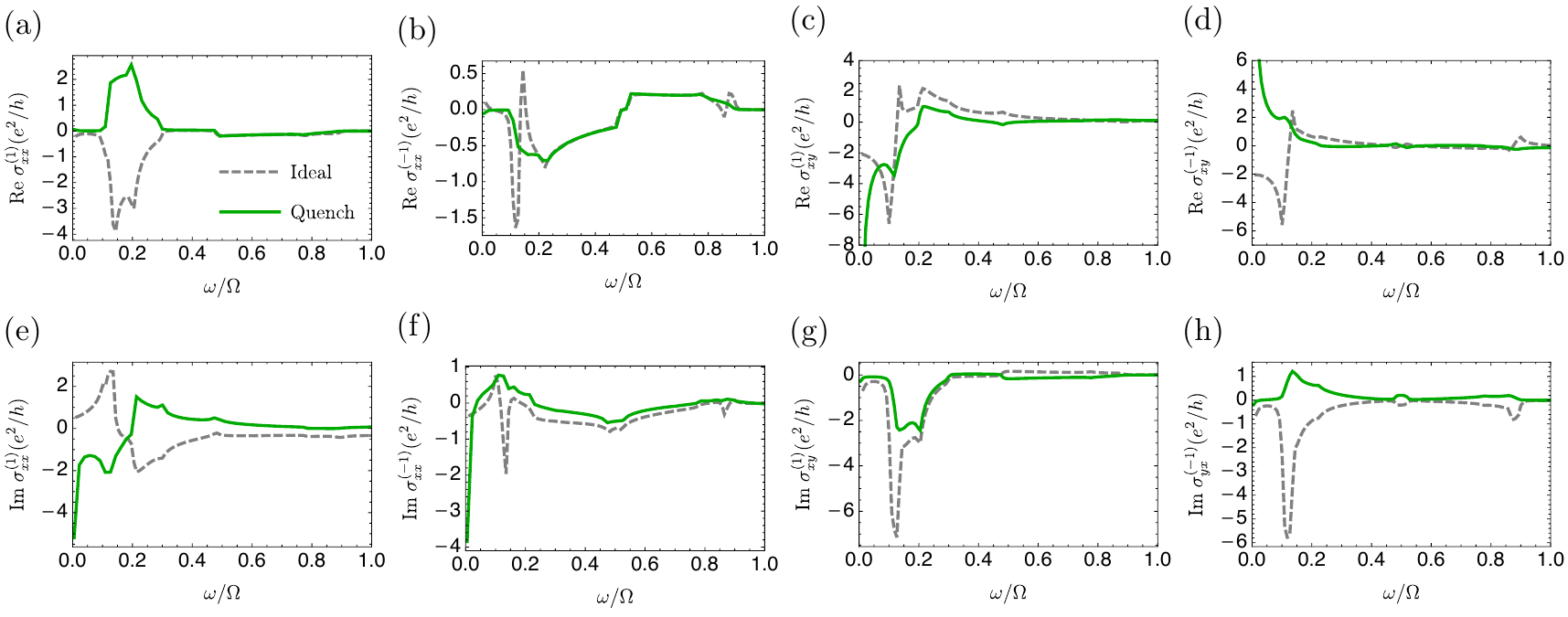}
   \caption{Heterodyne optical conductivity of the driven quantum well as a function of the probe frequency in the steady state following a quench by the drive (solid curve) compared to the ideal Floquet ocupation (dashed curve). The parameters are as in Fig.~\ref{fig:qw_2}(a-d).}
   \label{fig:qw_4}
\end{figure*}

\vspace{-3mm}
\subsubsection{Ideal Floquet occupation }

Figs.~\ref{fig:qw_1} and~\ref{fig:qw_2} summarize our numerical results for the ideal case when only the lower Floquet band is occupied. In Fig.~\ref{fig:qw_1}, we show the results for the high-frequency drive at $\Omega/A=8$. As before, the main features of the response correspond to optical transitions at the van Hove singularities of the Floquet bands. We note that in addition to steps and peaks at optical transitions $2\epsilon$ in the same Floquet zone, the heterodyne response ($n=\pm1$) also shows features at optical transitions $\mp(\Omega-2\epsilon)$ across neighboring Floquet zones. 

In the high-frequency regime, the homodyne Hall conductivity approaches a quantized value in the DC limit set by the Chern number of the Floquet band. Moreover, the heterodyne response, while nonzero, is suppressed by one or two orders of magnitudes. 

In Fig.~\ref{fig:qw_2}, we show the results for the mid-frequency drive at $\Omega/A=2.5$. We note that now the homodyne and heterodyne components have the same order of magnitude. At this frequency, the Floquet spectrum is modified by the resonant drive via gap closings at the Floquet zone edge. In particular, the Floquet bands have nontrivial Chern numbers $C=2$ both for $M/A=0.2$ and $C=1$ for $M/A=-0.2$. 

As before, the nontrivial topology of the Floquet bands is reflected in the DC limit of the homodyne Hall conductivity, consistent with general expectations. The DC limit of the heterodyne Hall conductivity shows an enhanced value. We study this limit based on Eq.~\ref{eq:sigmauv-DC} and show our results in Fig.~\ref{fig:heterodyne_3}. We observe that the DC heterodyne Hall conductivity is generically not quantized. However, both DC Hall and longitudinal heterodyne conductivity show singular enhancements at the topological transitions in the Floquet spectrum and can become large as the frequency is lowered. %

\subsubsection{Quench}

As an alternative dynamical scenario, we also compute optical response following a quench of the static Hamiltonian by the drive~\cite{Dehghani1,Yates,Du,Gavensky_2018}. We simplify the calculation by assuming that only the diagonal elements of the density matrix in the Floquet basis contribute. The contributions of the off-diagonal elements $g_{0\alpha\beta}$ are accompanied by phase factor $e^{-i(t-t_0)(\epsilon_\alpha-\epsilon_\beta)}$. So assuming $g_0$ does not depend strongly on the initial time, these contributions are expected to become incoherent for $t\gg t_0$. One can also see this by integrating over the initial distant time $t_0$, which formally cancels for all terms with $\epsilon_\alpha\neq\epsilon_\beta$. Thus, we take
\begin{equation}
g_{0 \alpha \beta} = | \inner{\phi_\alpha (0)}{\psi_\text{GS}} |^2 \delta_{\alpha \beta},
\end{equation}
where $\ket{\psi_\text{GS}}$ is the ground state of the quantum well without the drive.

A typical sample of our results in the mid-frequency regime, $\Omega/A=2.5$ and $M>0$, is shown in Fig.~\ref{fig:qw_3} for the homodyne response, and in Fig.~\ref{fig:qw_4} for the heterodyne response. The peaks and steps in the optical conductivity still follow the optical transition at van Hove singularities of the Floquet bands. However, in the quenched system, the intensities of these features is strongly modified, especially at lower optical frequencies. In particular, the DC Hall conductivity of the homodyne response is not quantized any more since it is now determined by the partial occupation of both bands, which carry opposite Berry fluxes.

\section{Conclusion}\label{sec:discuss}

In this work, we have extended the Kubo formula of the linear response to the case of a periodically driven system. Our expression is valid for a general density matrix of the driven system. These expressions simplify when the density matrix is diagonal in the Floquet basis of quasienergies or when only the diagonal part is taken to contribute to the response, upon averaging the initial time of switching on the probe.

We have derived the Floquet optical conductivity and elucidate its general dynamical structure and, in particular, its homodyne and heterodyne components. Importantly, the non-homodyne response means that when the system is probed at some frequency $\omega$, a current is generated not only at the drive frequency but also at frequencies $\omega+n\Omega$, where $\Omega$ is the drive frequency and $n$ is an integer. We also obtain Floquet-optical sum rules which include an inverse effective mass term related to the curvature of the Floquet.

Using the general expressions for the optical conductivity and resonant and off-resonant rotating-wave approximation, we obtain analytical results for the case of a driven two-level system in the high-frequency limit when the drive doesn't necessarily commute with the system's Hamiltonian.  We investigate the optical response of two driven lattice models: the Haldane model on the honeycomb lattice with circularly polarized light and the BHZ model for a two dimensional quantum well topological insulator with an oscillatory Zeeman field. In both models we calculate numerically the longitudinal and Hall conductivity response to an AC probe field, including the homodyne and heterodyne components. We observe that the steps and peaks of the optical conductivity trace the optical transitions of the quasienergy spectrum at van Hove singularities. 

The homodyne and heterodyne responses trace optical transitions, respectively, within and across neighboring Floquet zones.  Moreover, the DC limit of homodyne and heterodyne conductivities signal the Floquet topological transitions. In the ideal Floquet occupation, the DC homodyne Hall conductivity is quantized at the Chern number of the occupied Floquet bands. The DC heterodyne response, on the other hand, shows singular enhancement at the Floquet phase transitions. Away from the ideal limit, the quantization is spoiled; however, the spectral signatures in the Floquet optical response persist. These features demonstrate that the full Floquet optical conductivity is a powerful probe of the Floquet spectrum and its nontrivial topology.

\begin{acknowledgments}
We acknowledge useful conversations with Takashi Oka and initial contributions by Aaron Farrell and Aaron McClendon in the early stages of this work.  This work is supported in part by the National Science Foundation CAREER grant DMR-1350663, the Binational
Science Foundation grant No. 2014345, the College of Arts and Sciences at Indiana University, NSERC and FRQNT. M.R-V. was primarily supported by the National Science Foundation through the Center for Dynamics and Control of Materials: an NSF MRSEC under Cooperative Agreement No. DMR-1720595. We acknowledge the hospitality of Aspen Center for Physics, supported by National Science Foundation grant PHY-1607611, where parts of this work were performed. B.S. and M.R.V thank the hospitality and funding support of the Max-Planck Institute for the Physics of Complex Systems in Dresden, Germany, where parts of this work were performed.
\end{acknowledgments}

\appendix*
\begin{widetext}
\section{DC limit}
In this Appendix we provide some details of the derivation of the DC limit of the optical Hall conductivity, $\lim_{\omega\to0}\bbsigma_{xy}^{(n)}(\omega)$. In particular, we show that the divergent $1/\omega$ terms vanish for all components of the optical Hall conductivity. These terms are found from Eq.~(\ref{eq:sigman_diag}) by setting $\omega=0$ inside the bracket,
\begin{equation}\label{eq:DCsigman-div}
\bbsigma^{uv(n)}_\text{div}(0) = \frac i\omega \sum_{\alpha}g_{0\alpha}\left[\sum_{\gamma m}\left( 
\frac{
j^{u(n+m)}_{0\alpha\gamma}j^{v(-m)}_{0\gamma\alpha}
}
{\epsilon_\alpha-\epsilon_\gamma-m\Omega + i0^+}
+
\frac{
j^{v(m)}_{0\alpha\gamma}j^{u(n-m)}_{0\gamma\alpha}
}
{\epsilon_\alpha-\epsilon_\gamma-m\Omega - i0^+}
\right) + \mathbb{m}^{uv(n)}_{0\alpha\alpha}\right].
\end{equation}

First, focus on the terms in parentheses. Using the identity $(z+i0^+)^{-1} = \mathcal{P}(z^{-1}) - i\pi\delta(z)$, we see that the terms with the delta function boil down to 
\begin{equation}
    \frac{1}{4\pi\omega}\sum_{\alpha\neq\gamma,m}\int \left[
    j^{u(n+m)}_{0\alpha\gamma}(\vex k) j^{v(-m)}_{0\gamma\alpha}(\vex k)
    -
    j^{u(m)}_{0\alpha\gamma}(\vex k) j^{v(n-m)}_{0\gamma\alpha}(\vex k)
    \right]
    \delta(\epsilon_\alpha(\vex k)-\epsilon_\gamma(\vex k) - m\Omega) d\vex k,
\end{equation}
where we have shown the dependence on the crystal momentum $\vex k$ explicitly for clarity, so $\alpha$ and $\gamma$ here refer to band and other internal indices. We fix the values of the quasienergies in the first Floquet zone, so the delta function enforces $m=0$. However, in the simplest case we are considering here, the bands are nondegenerate for a given value of $\vex k$, so the condition $\epsilon_\alpha=\epsilon_\gamma$ cannot be satisfied and these terms vanish.

The other terms read
\begin{align}
    \frac i\omega \mathcal{P} \sum_{\alpha\neq\gamma,m} g_{0\alpha} \frac{
    j^{u(n+m)}_{0\alpha\gamma}j^{v(-m)}_{0\gamma\alpha}
    +
    j^{v(m)}_{0\alpha\gamma}j^{u(n-m)}_{0\gamma\alpha}
    }
    {\epsilon_\alpha-\epsilon_\gamma-m\Omega}
    =
    \frac 1\omega \sum_{\alpha\neq\gamma,m}g_{0\alpha}
    \left[
    j^{u(n+m)}_{0\alpha\gamma}r^{v(-m)}_{\gamma\alpha}
    -
    r^{v(m)}_{\alpha\gamma}j^{u(n-m)}_{0\gamma\alpha}
    \right],
\end{align}
\end{widetext}
where we have used Eq.~(\ref{eq:jmr}) to relate to the elements of the Berry connection. The bracket on the right-hand side here vanishes for $\gamma=\alpha$ identically when summed over $m$, so we can sum over all $\gamma$ and $m$ without singularities, yielding
\begin{equation}
\frac1{\omega}\sum_{\alpha} g_{0\alpha}\left[ j^u_0(t) r^v - r^v j^u_0(t) \right]^{(n)}_{\alpha\alpha}.
\end{equation}
Since $j^u(t) = \partial H(t)/\partial_{k_u}$ and $r^v = i\partial_{k_v}$, the terms in the bracket evaluate to $-i\partial^2 H(t)/\partial k_u\partial k_v \equiv -i\mathbb{m}_0^{uv}(t)$. So, this term simplifies to
\begin{equation}
    -\frac{i}{\omega}\sum_{\alpha}g_{0\alpha} \mathbb{m}^{uv(n)}_{0\alpha\alpha},
\end{equation}
which exactly cancels the divergent $\mathbb{m}$ term in Eq.~(\ref{eq:DCsigman-div}). So, all divergent terms vanish in the DC limit.

The next order in $\omega$ yields the finite DC limit as
\begin{widetext}
\begin{align}
    \bbsigma^{uv(n)}(0) &= 
    -i \sum_{\gamma\neq\alpha,m} g_{0\alpha} \frac{
    j^{u(n+m)}_{0\alpha\gamma}j^{v(-m)}_{0\gamma\alpha}
    -
    j^{v(m)}_{0\alpha\gamma}j^{u(n-m)}_{0\gamma\alpha}
    }
    {(\epsilon_\alpha-\epsilon_\gamma-m\Omega)^2} \label{eq:sigmauv-DC}\\ 
    &=
    -i \sum_{\gamma\neq\alpha,m} g_{0\alpha} \left[
    r^{u(n+m)}_{\alpha\gamma}r^{v(-m)}_{\gamma\alpha}
    -
    r^{v(m)}_{\alpha\gamma}r^{u(n-m)}_{\gamma\alpha}
    \right] 
    +
    in\Omega
    \sum_{\gamma\neq\alpha,m} g_{0\alpha} \frac{
    r^{u(n+m)}_{\alpha\gamma}r^{v(-m)}_{\gamma\alpha}
    +
    r^{v(m)}_{\alpha\gamma}r^{u(n-m)}_{\gamma\alpha}
    }
    {\epsilon_\alpha-\epsilon_\gamma-m\Omega},
\end{align}
where we have again used Eq.~(\ref{eq:jmr}) to relate to the elements of the Berry connection. For Hall conductivity ($uv=xy$), as shown under Eq.~(\ref{eq:DCHall0}), the first term is $\frac1{2\pi}\sum_{\alpha} g_{0\alpha} C_\alpha \delta_{n0}$. The second term is as shown in Eq.~(\ref{eq:DCHalln-rr}). To obtain the form given in Eq.~(\ref{eq:DCHalln-G}), we first note again that the term $\gamma=\alpha$ would vanish upon integration over $m$ since its summand is odd under $m$, so can sum over all $\gamma$. Now, writing matrix elements $r^{v(m)}_{\alpha\gamma} = \bbra{\phi_{\alpha 0}}r^v\kket{\phi_{\gamma m}}$ and using the definition of the Floquet Green's function~(\ref{eq:FG}), we can write this term as
\begin{align}
    &~~~~ in\Omega \sum_{\alpha}g_{0\alpha} \left[\bbra{\phi_{\alpha -n}}
    r^u \hhat{G}(\epsilon_\alpha) r^v
    \kket{\phi_{\alpha 0}}
    +
    \bbra{\phi_{\alpha 0}}
    r^v \hhat{G}(\epsilon_\alpha) r^u
    \kket{\phi_{\alpha n}}\right] \nonumber \\ &=
    in\Omega \sum_{\alpha}g_{0\alpha}
    \left[
    \bbra{\phi_{\alpha -n}}
    [r^u , \hhat{G}(\epsilon_\alpha)]
    r^v
    \kket{\phi_{\alpha 0}}
    +\frac1{n\Omega}
    \bbra{\phi_{\alpha -n}}
    r^u r^v
    \kket{\phi_{\alpha 0}}
    \right. 
    + \left.
    \bbra{\phi_{\alpha 0}}
    r^v [\hhat{G}(\epsilon_\alpha), r^u]
    \kket{\phi_{\alpha n}}
    -\frac1{n\Omega}
    \bbra{\phi_{\alpha 0}}
    r^v r^u
    \kket{\phi_{\alpha n}}
    \right] \nonumber\\ &=
    in\Omega \sum_\alpha g_{0\alpha}
    \bbra{\phi_{\alpha 0}}
    [r^u,\hhat G(\epsilon_\alpha+n\Omega)]r^v-r^v[r^u,\hhat G(\epsilon_\alpha)]
    +\frac1{n\Omega}
    [r^u, r^v]
    \kket{\phi_{\alpha n}} \nonumber\\ &=
    in\Omega\sum_\alpha g_{0\alpha}
    \bbra{\phi_{\alpha 0}}
    [[r^u,\hhat G^+_{\alpha n}],r^v] + \{[r^u,\hhat G^-_{\alpha n}],r^v\}
    \kket{\phi_{\alpha n}} \nonumber\\ &=
    in\Omega \sum_\alpha g_{0\alpha}
    \bbra{\phi_{\alpha 0}}
    \partial^2 \hhat G^+_{\alpha n}/\partial{k_v}\partial{k_u} + \{ \partial G^-_{\alpha n}/\partial{k_u}, \partial/\partial{k_v} \}
    \kket{\phi_{\alpha n}}.
\end{align}
\end{widetext}
In the first and second lines, we have used the identities $\hhat{G}(\omega)\kket{\phi_{\alpha n}} = (\omega - \epsilon_\alpha - n\Omega)^{-1}\kket{\phi_{\alpha n}}$ and $\bbra{\chi_{m}}\hhat G(\omega)\kket{\psi_{n}} = \bbra{\chi_{m-n}}\hhat G(\omega-n\Omega)\kket{\psi_0}$. In the next lines we used the identities have used $[r^u,r^v]=0$ and $[[r^u,\hhat G],r^v] = \partial_{k_v}\partial_{k_u} \hhat{G}$, and defined $\hhat G^\pm_{\alpha n} = \frac12 [\hhat G(\epsilon_\alpha+n\Omega)\pm \hhat G(\epsilon_\alpha)]$.

\bibliographystyle{apsrev}

\end{document}